\documentclass[12pt,preprint]{aastex}

\newcommand{\be}{\begin{equation}}
\newcommand{\ee}{\end{equation}}
\newcommand{\kkon}{{ {\kappa} }}
\newcommand{\const}{ {\Gamma} } 
\newcommand{\uvec}{{ {\bf u} }}
\newcommand{\bvec}{{ {\bf B} }}
\newcommand{\zhat}{{ {\hat z} }} 
\newcommand{\xhat}{{ {\hat x} }} 
\newcommand{\rhat}{{ {\hat r} }} 
\newcommand{\phat}{{ {\hat p} }} 
\newcommand{\thetat}{{ {\hat \theta} }} 
\newcommand{\phihat}{{ {\hat \phi} }}
\newcommand{\epj}{{ {\underline{\epsilon}}_j}} 
\newcommand{\epp}{{ {\underline{\epsilon}}_p}} 
\newcommand{\epq}{{ {\underline{\epsilon}}_q}} 
\newcommand{\epphi}{{ {\underline{\epsilon}}_\phi}} 
\newcommand{\vecomega}{{ {\vec \Omega} }} 
\newcommand{\rotcon}{{ \omega }} 
\newcommand{\divnum}{{ \Upsilon }} 
\newcommand{\roterm}{{ \Lambda }}  
\newcommand{\monopole}{{ {\cal M} }} 
\newcommand{\pcomp}{{ X_{\rm P} }} 
\newcommand{\qcomp}{{ X_{\rm Q} }} 
\newcommand{\lambar}{{ \langle \lambda \rangle }} 
\def\lta{\,\raise 0.3 ex\hbox{$ < $}\kern -0.75 em
 \lower 0.7 ex\hbox{$\sim$}\,}
\def\gta{\,\raise 0.3 ex\hbox{$ > $}\kern -0.75 em
 \lower 0.7 ex\hbox{$\sim$}\,} 

\begin{document} 

\title{\bf MAGNETICALLY CONTROLLED ACCRETION FLOWS 
ONTO YOUNG STELLAR OBJECTS} 

\author{Fred C. Adams$^{1,2}$ and Scott G. Gregory$^{3}$} 

\affil{$^1$Michigan Center for Theoretical Physics \\
Physics Department, University of Michigan, Ann Arbor, MI 48109} 

\affil{$^2$Astronomy Department, University of Michigan, Ann Arbor, MI 48109} 

\affil{$^3$Department of Astrophysics, California Institute of
  Technology, Pasadena, CA 91125}

\begin{abstract} 

Accretion from disks onto young stars is thought to follow magnetic
field lines from the inner disk edge to the stellar surface. The
accretion flow thus depends on the geometry of the magnetic field.
This paper extends previous work by constructing a collection of
orthogonal coordinate systems, including the corresponding
differential operators, where one coordinate traces the magnetic field
lines. This formalism allows for an (essentially) analytic description
of the geometry and the conditions required for the flow to pass
through sonic points. Using this approach, we revisit the problem of
magnetically controlled accretion flow in a dipole geometry, and then
generalize the treatment to consider magnetic fields with multiple
components, including dipole, octupole, and split monopole
contributions.  This approach can be generalized further to consider
more complex magnetic field configurations. Observations indicate that
accreting young stars have substantial dipole and octupole components,
and that accretion flow is transonic. If the effective equation of
state for the fluid is too stiff, however, the flow cannot pass
smoothly through the sonic points in steady state. For a multipole
field of order $\ell$, we derive a general constraint on the
polytropic index, $n > \ell$ + 3/2, required for steady transonic flow
to reach free-fall velocities.  For octupole fields, inferred on
surfaces of T Tauri stars, the index $n > 9/2$, so that the flow must
be close to isothermal. The inclusion of octupole field components
produces higher densities at the stellar surface and smaller areas for
the hot spots, which occur at higher latitudes; the magnetic
truncation radius is smaller (larger) for octupole components that are
aligned (anti-aligned) with the stellar dipole. This contribution thus
increases our understanding of magnetically controlled accretion for
young stellar objects and can be applied to a variety of additional
astrophysical problems.

\end{abstract} 


\section{Introduction}
\label{sec:intro}

During the star formation process, most of the mass that becomes part
of the nascent star initially falls onto a surrounding circumstellar
disk, rather than directly onto the stellar surface.  This accretion
process can be (indirectly) observed during the T Tauri phase of
evolution, after the star/disk system has emerged from its
protostellar envelope.  The transfer of material from the disk to the
star takes place at the inner disk edge, which generally does not
extend to the stellar surface.  Instead, the inner edge is connected
to the star through magnetic fields, and accretion takes place along
these field lines.

This paradigm of magnetic accretion was developed in the early 1990s
for classical T Tauri stars (e.g., \citealt{kon91,shu94,har94}) and is
roughly analogous to that of magnetically controlled accretion from
disks onto neutron stars \citep{gho78} and black holes (Blanford \&
Payne 1982; see Uzdensky 2005 for further discussion).  Many of the
diverse observational characteristics of accreting T Tauri stars can
be explained within the basic framework of the magnetospheric
accretion scenario.  The shapes of spectral energy distributions
(SEDs) in the near infrared are consistent with cavities in inner dust
disks (e.g., \citealt{ken87,als88,rob07}), although gas likely extends
closer to the star \citep{naj03}.  The excess emission, primarily at
IR and UV wavelengths and apparent from the SEDs of accreting T Tauri
stars, can be explained by the reprocessing of the stellar photons by
dusty material in circumstellar disks and from shock emission at the
base of the accretion columns. Gas in the accretion flow rains down
onto the stellar surface producing hotspots that radiate primarily in
the UV and also in the soft X-ray waveband (\citealt{kas02,arg11}).
Accretion related hotspots, in addition to cool spots which arise
where bundles of magnetic flux rise through the stellar surface into
the atmosphere, contribute to the high level of photometric
variability of T Tauri stars \citep{bou93}.  Spectroscopically, the
photospheric absorption lines of accreting T Tauri stars are typically
shallower than those observed from non-accreting T Tauri stars and
main-sequence stars of the same spectral type (see Figure 1 of
\citealt{har91}), as a result of the additional continuum emission
from the accretion spots \citep{cal98}.  Many emission lines often
exhibit red-shifted and blue-shifted absorption components, sometimes
simultaneously, characteristic of accretion and outflows, respectively
(\citealt{edw94,fis08}), and suggesting that the star-disk interaction
region contains complex kinematic gas flows with material both
accreting onto the star and being launched from the system in outflows
(see \citealt{bou07} for a review).

A key assumption of the magnetospheric accretion model is that T Tauri
stars support large-scale magnetic fields that are sufficiently
globally ordered and strong enough to disrupt the disk at a distance
of few stellar radii.  As demonstrated by \citet{kon91}, a stellar
dipole field of polar strength $B\sim10^3$ G is sufficient. Initial
magnetospheric accretion models focused on dipole magnetic fields (see
also \citealt{li96, li99}), and have included detailed heating and
cooling calculations \citep{mar96}, polytropic equations of state
\citep{kol02}, detailed numerical treatments \citep{rom02,zan09}, and
dipole fields that are tilted with respect to the stellar rotation
axis \citep{rom03}.

Strong stellar-disk-averaged surface fields have now been measured on
a number of T Tauri stars in different star forming regions, most
successfully through the detailed analysis of magnetically sensitive,
and therefore Zeeman broadened, lines in intensity spectra
\citep{joh07,yan11}.  However, such broadening measurements give no
information about the stellar magnetic field topology. Another
manifestation of the Zeeman effect, namely the circular polarization
of magnetically sensitive lines, does yield information about the
field geometry.  Initial spectropolarimetric studies show mixed
results: Attempts at measuring the polarization signal in photospheric
absorption lines often fail to detect the presence of surface magnetic
fields (e.g., \citealt{joh86}).  However, as opposite polarity
(positive and negative) surface field regions give rise to signals
that are polarized in the opposite sense, a net circular polarization
signal of zero is consistent with T Tauri stars hosting complex
surface fields \citep{val04}.  In contrast, a strong and rotationally
modulated circular polarization signal was detected in the HeI D$_3$
(5876{\AA }) emission line by \citet{joh99}.  This particular line of
helium has a high excitation potential and is thought to form at the
base of accretion columns.  The strong rotationally modulated signal
in this, and other accretion related emission lines, is found to be
well described by a simple model where the bulk of the accreting gas
lands on the stellar surface in a single polarity radial field
spot \citep{val04}.  This finding suggests that even though accreting
T Tauri stars host complex surface magnetic fields, their large-scale
field topology, and in particular the portion of the field that
carries gas from the inner disk to the star, is simpler and globally
well-ordered.

The non-dipolar nature of T Tauri magnetic fields has recently been
confirmed.  Spectropolarimetric observations, combined with
tomographic imaging techniques whereby the rotational modulation of
the Zeeman signal is modeled, have allowed magnetic surface maps to be
derived for a number of accreting T Tauri stars
\citep{don07,don08,hus09,don10v2247,don10aatau,don11v2129,don11twhya,don11v4046}.
For completeness we note that magnetic maps have also been published
for a handful of non-accreting T Tauri stars
\citep{dunst08,ske10,mars10,wait11}.  These maps are constructed by
considering the rotational modulation of the polarization signal
detected in both photospheric absorption lines, which form uniformly
across the entire stellar surface, and the signal in accretion related
emission lines, which trace the field at the base of accretion
columns. In practice, the polarization signals detected in
photospheric absorption lines are small, and cross-correlation
techniques (e.g., \citealt{don97}) are employed in order to extract
information from as many spectral lines as possible (see
\citealt{don09} for a review of Zeeman-Doppler imaging, the technique
used to construct stellar magnetic maps, and \citet{don10aatau} for
its specific application to T Tauri stars).
    
The observationally derived magnetic maps can be decomposed into the
various spherical harmonic modes.  Some stars are found to host very
complex magnetic fields with many high order components and strong
toroidal field components, such as both stars of the close binary
V4046~Sgr \citep{don11v4046}, V2247~Oph \citep{don10v2247}, CR~Cha and
CV~Cha \citep{hus09}.  Intriguingly, however, many accreting T Tauri
stars have field topologies that are well described as dipole-octupole
composite fields.  The polar strength of the dipole and octupole field
component varies from star to star.  AA~Tau has a dominantly dipolar
magnetic field, with a weak octupole component \citep{don10aatau},
whereas TW~Hya hosts a dominantly octupolar magnetic field with a weak
dipole component \citep{don11twhya}.  The same is true for V2129~Oph,
although the dipole component has been observed to vary by a factor of
three (from a polar strength of $\sim0.3$ to $\sim0.9\,{\rm kG}$) over
a timescale of four years perhaps hinting at the existence of a
magnetic cycle \citep{don11v2129}.  BP~Tau, one of the best studied
accreting T Tauri stars, hosts a magnetic field with both strong
dipole and octupole field components \citep{don08}.

These observations provide clear motivation to consider magnetic
fields with both dipole and octupole components in models of accretion
flow, and motivated (at least in part) by the availability of the new
observational data, magnetospheric accretion models with higher order
multipole stellar magnetic fields have been developed
\citep{gre06,lon07,lon08,moh08,gre10}.  To date, flow models with
accretion taking place along dipole stellar field lines have
successfully reproduced emission line profiles and their rotational
variability, including helium lines \citep{ber01,fis08,kur11}, calcium
lines \citep{aze06} and hydrogen lines
\citep{har94,muz01,sym05,lim10,kur11}.  However, the simulated line
profiles based on dipole stellar magnetospheres show more variability
than is observed \citep{sym05}, and for some stars the inferred
trajectory of accretion flows close to the stellar surface, where the
higher order field components will influence the infalling columns of
gas \citep{gre08}, are inconsistent with the dipole flow model
\citep{fis08}.

Deep, broad, and often rotationally modulated red-shifted absorption
components are commonly detected in accretion related emission lines
(e.g., \citealt{edw94,bou03,edw06,fis08}).  Analysis of such features
allows the determination of the kinematic gas flow (the accretion
column) crossing the line-of-sight to the star. The material arrives
at the stellar surface with highly supersonic speeds of several
hundred km s$^{-1}$.  Since the inflowing material leaves the inner
disk edge at low (subsonic) speeds, the gas must make a sonic
transition while it follows magnetic field lines from the disk onto
the star. As a result, accretion is described by transonic flow
solutions (which are essentially the reverse of the well-known Parker
model of the Solar wind; see \citealt{par65}). As we show in this
paper, the requirement of a smooth transition through the sonic point,
the need for free-fall speeds in the inner limit, and the divergence
properties of higher order multipole moments, jointly place strong 
constraints on the allowed polytropic index of the accretion flow.

\begin{figure} 
\figurenum{1} 
{\centerline{\epsscale{0.90} \plotone{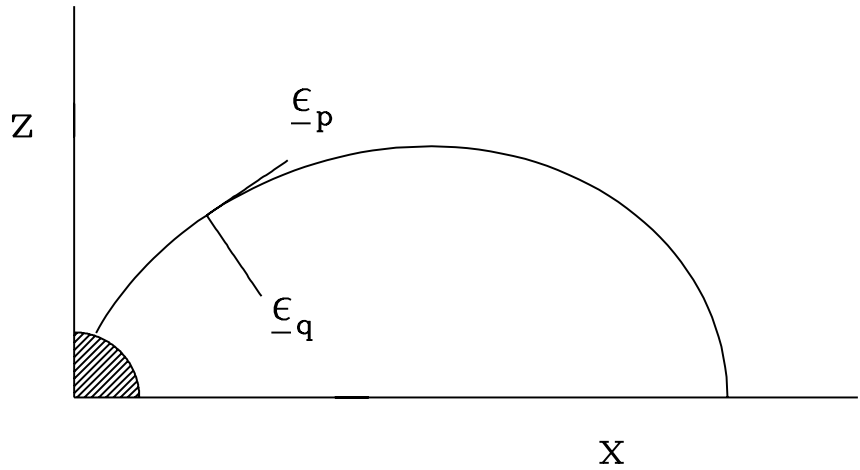} } } 
\figcaption{Schematic diagram showing the coordinates used here, 
for one quadrant of the poloidal plane.  The star is shown in the
lower left corner. The curve depicts one magnetic field line that
connects the stellar surface to the inner edge of the disk. At each
point along the field line, basis vectors describe the coordinate
system. At one such point, the basis vectors are shown.  The field
lines correspond to lines of constant $q$, so the basis vector $\epq$
= $\nabla q$ is perpendicular to the field line. The coordinate $p$
provides a measure of the distance along the field line, so that the
basis vector $\epp$ = $\nabla p$ points along the field line. }
\label{fig:diagram} 
\end{figure}

The goal of this paper is relatively modest: Building on the
theoretical work outlined above, this paper provides an analytic, or
at least semi-analytic, treatment of magnetically controlled accretion
flows (where the term ``semi-analytic'' refers to models where the
equations are reduced to, at most, ordinary differential equations).
Motivated by the observational work outlined above, we focus on
transonic solutions and generalize existing work to include higher
order multipoles, especially magnetic fields with both dipole and
octupole components.  To reach this objective, we construct (novel)
orthogonal coordinate systems for each given magnetic field
configuration under consideration. One of the coordinates (denoted
here as $p$) follows the magnetic field lines, whereas the other
coordinate (denoted here as $q$) is orthogonal to the first in the
poloidal plane (see Figure \ref{fig:diagram}).  We specialize to the
case of axial symmetry, which applies to systems where the magnetic
field configuration co-rotates with the central star, and where the
field is strong (so that the toroidal field component is small and the
angular velocity along a streamline is constant).

Because this paper constructs coordinate systems $(p,q)$ in the
poloidal plane, it is important to outline why this framework is
useful: The coordinate $p$ follows the magnetic field lines, and hence
the streamlines, so that the fluid fields are functions of the
coordinate $p$, and the value of $p$ measures the position along the
field line. By construction, $\nabla p$ is parallel to the magnetic
field. As we show below, the divergence operator that describes the
flow includes the quantity $|\nabla p|$, which is the inverse of one
scale factor of the coordinate system (e.g., Weinreich 1998). This
scale factor must be included to properly describe the divergence and
hence the physics of the flow. The perpendicular coordinate $q$ labels
the streamlines, i.e., the flow follows lines of constant $q$.  The
flow trajectory is thus specified by the functions $\xi(\theta)$ or
equivalently $\theta(\xi)$ along a streamline ($\xi$ and $\theta$ are
spherical coordinates); these relations are necessary to evaluate the
relevant functions along the field lines. As a result, one must have a
description of the scalar fields $p$ and $q$ that make up the
coordinate system in order to solve for the flow. 
However, one could, in principle, include the correct scale factor
($h_p=|\nabla{p}|^{-1}$) and use the functions $\xi(\theta)$ that
specify streamlines (lines of constant $q$) without an explicit
reference to the fact that $p$ and $q$ are coordinates.  This strategy
has been used previously for dipole fields (Blandford \& Payne 1982,
Hartmann et al. 1994, and others).  For more complicated magnetic
field configurations, considered here, it is more straightforward to
define the coordinate system and work within it --- we gain additional
physical understanding by being aware of the coordinate system in
which the flow is taking place (for dipole fields, such coordinates
have been used to study accretion onto white dwarfs -- see Canalle et
al. 2005, Saxton et al. 2007).  Finally, the resulting framework can
be readily generalized for more complex magnetic fields (Appendix
\ref{sec:mathappendix}) and can be used in a variety of other
applications (e.g., outflows from Hot Jupiters, Adams 2011).

This paper is organized as follows.  We start with a general
discussion of coordinate systems in Section \ref{sec:coord}. The
equations of motion for fluid flow are then considered in Section
\ref{sec:fluid}, where they are formulated in terms of these new
coordinate systems; by construction, the fluid fields are functions of
only one variable, which measures the location along the magnetic
field line(s). We then derive a general constraint on the polytropic
index $n$ required for the flow to reach free-fall speeds; since $n$
must be large, the flow can be described to good approximation using
an isothermal equation of state.  In Section \ref{sec:dipole}, we
consider the magnetic field to be that of a pure dipole; although this
case has been studied previously, we re-formulate the problem in new
coordinates to illustrate our approach.  In Section \ref{sec:dipoct},
we consider more complicated magnetic field configurations including
both dipole and octupole components (consistent with current
observations of accreting young stars).  The inner disk edge is
truncated magnetically, and hence the inner boundary depends on the
magnetic field structure. Using the magnetic field configurations of
this paper, we revisit the magnetic truncation radius in Section
\ref{sec:truncate}. The paper concludes, in Section
\ref{sec:conclude}, with a summary and discussion of our results.
During the course of this work, we have derived a number of
mathematical results that apply to general coordinate systems of the
form considered herein; these results are collected in Appendix
\ref{sec:mathappendix}. For completeness, we also consider magnetic
fields with dipole and radial (split-monople) components (Appendix
\ref{sec:diprad}); this configuration arises when the infall-collapse
flow that forms the disk drags in magnetic field lines from the
original molecular cloud core, and also when stellar winds open up
field lines to become (nearly) radial. Finally, Appendix
\ref{sec:conservemass} provides a consistency check by showing how
this formalism explicitly conserves mass.

\newpage 
 
\section{Construction of Coordinate Systems} 
\label{sec:coord}

In this treatment, we develop a series of coordinate systems that
follow the magnetic field lines. We thus start with a given magnetic
field configuration ${\bf B} (r,\theta)$, which we take to be
axisymmetric (the magnetic field is poloidal, with no toroidal
component, $B_{\phi}$ = 0).  The origin is located at the center of
the star. Since the star is expected to be rotating, the coordinate
system co-rotates with the star, and is thus fixed. The effects of
rotation are then incorporated by including (the usual) non-inertial
terms in the equations of motion (see Section \ref{sec:fluid}).  By
definition, coordinate systems are defined by scalar fields, which we
denote here as $(p,q)$. If the magnetic field is curl-free (and hence
current-free), then the field ${\bf B}$ can be written as the gradient
of a scalar field.  The coordinate $p$ specifies the distance along
the field line and hence the gradient $\nabla p$ points in the
direction of the field line. The coordinate $q$ is perpendicular so
that $\nabla q \cdot \nabla p$ = 0.  Notice that lines of constant $q$
correspond to the magnetic field lines; similarly, lines of constant
$p$ correspond to equipotentials of the multipole field under
consideration.  The coordinates $(p,q)$ are thus orthogonal
coordinates in the poloidal plane, and can be used instead of the more
familiar spherical coordinates $(r,\theta)$.  The azimuthal angle
$\phi$ is the same for both cases and provides the third orthogonal
coordinate. We could also use cartesian coordinates, where the pair
$(x,z)$ define the poloidal plane.

As outlined above, this treatment neglects the azimuthal field
component. In some systems, the presence of a twisted field component
($B_\phi \ne 0$) above the disk is necessary for maintaining balance
in the angular momentum transport (Long et al. 2005).  However, recent
numerical simulations (e.g., Romanova et al. 2011, Long et al. 2011)
indicate that the initial potential field structure is retained for
regions within the disk truncation radius, so that our approximation
($B_\phi=0$) is viable. For completeness we note that, in general, the
field will not strictly be a potential field when the azimuthal
component is nonzero (although a potential field provides a good
approximation); the azimuthal component also affects angular momentum
transport.

Specification of a coordinate system requires not only the coordinates
themselves, but also the basis vectors. For orthogonal coordinate
systems, one can also use the scale factors.  The set of covariant
basis vectors $\epj$ arises from the gradients of the scalar fields
that define the coordinates (Weinreich 1998), in this case 
$\epp \equiv \nabla p$ and $\epq \equiv \nabla q$.  Note that these
quantities are basis vectors, rather than unit vectors, so that their
length is not, in general, equal to unity. The corresponding scale
factors $h_j$ are given by the relation
\be
h_j = \left| \epj \right|^{-1} \, .
\label{hfactordef} 
\ee
The schematic diagram of Figure \ref{fig:diagram} depicts one magnetic
field line and the basis vectors at one point along the field line.
The coordinate $p$ measures the distance along the field line and the
basis vector $\epp = \nabla p$ points in the direction of the field
line. The coordinate $q$ is constant along field lines, and thus
labels both the field lines and the streamlines. The basis vector
$\epq = \nabla q$ is perpendicular to the field line.

In magnetically controlled accretion, one important geometrical effect
is that the divergence operator takes a non-standard form.  For
orthogonal coordinate systems, the general form of the divergence
operator is given by
\be
\nabla \cdot {\bf V} = {1 \over h_p h_q h_\phi} 
\left[ {\partial \over \partial p} \left( h_q h_\phi V_p \right) + 
{\partial \over \partial q} \left( h_p h_\phi V_q \right) \right] + 
{1 \over \xi \sin\theta} {\partial V_\phi \over \partial \phi} \, . 
\ee
The quantities $h_p$, $h_q$, and $h_\phi$ are the scale factors of the
coordinate system, as defined by equation (\ref{hfactordef}). 

In this application, we consider the fields to be axisymmetric so
that the $\phi$ derivatives vanish. Further, for flow along field
lines, the vector fields (e.g., the velocity field) have only one
component and depend on only one coordinate, so that the divergence
operator collapses to the form 
\be
\nabla \cdot {\bf V} = {1 \over h_p h_q h_\phi} 
{\partial \over \partial p} \left( h_q h_\phi V_p \right) =
{1 \over h_p} {\partial V_p \over \partial p} + 
{V_p \over h_p h_q h_\phi} {\partial \over \partial p} 
\left( h_q h_\phi \right) \, . 
\label{divsimple} 
\ee
As shown in Appendix \ref{sec:mathappendix}, $h_q h_\phi$ = $h_p$ and
$h_p \propto |{\bf B}|^{-1}$ (see {\bf Result 1} and the text below).  
For flow along the field lines, the divergence operator
(\ref{divsimple}) thus has a form consistent with the well-known
result that the effective area element of a moving fluid element
varies inversely with the field strength (Parks 2004).

\section{Fluid Equations for Magnetically Controlled Flow} 
\label{sec:fluid} 

In this treatment, we consider steady-state solutions and assume that
the magnetic field structure is strong enough to dominate the flow. As
a result, the magnetic field is held fixed and fluid elements must
follow the field lines.  The magnetic field is assumed to rotate with
the star, so we work in a rotating reference frame with rotation rate
${\vecomega} = \Omega \zhat$. Under these approximations, the
equations of continuity, force, and induction reduce to the forms
\be
\nabla \cdot \left( \rho \uvec \right) = 0 \, , \qquad 
\uvec\cdot \nabla \uvec + \nabla \Psi + 
{\vecomega} \times ({\vecomega} \times {\bf r}) + 
{1\over \rho} \nabla P = 0 \, , 
\qquad {\rm and} \qquad \bvec = \kkon \, \rho \, \uvec \, , 
\label{simple} 
\ee
where the parameter $\kkon$ is constant along streamlines. Note that
the magnetic field does not contribute a force along the field lines
and that the Coriolis force also does not contribute. The correction
to the effective gravity due to the rotating reference frame is
included.

The velocity vector ${\bf u}$ follows the magnetic field lines.  In
the following sections, we construct coordinate systems such that one
of the coordinates (denoted here as $p$) follows the field lines.  As
a result, the flow velocity has only one component, which points in
the direction of the magnetic field $\phat$ = $h_p \, \epp$.  We also
assume that the thermodynamics of the flow can be modeled using a
polytropic equation of state of the general form
\be
P = K \rho^{1 + 1/n} \, , 
\label{eqostate} 
\ee
where $n$ is the polytropic index. As shown below, however, the 
index $n$ must be relatively large in order for the flow to pass 
smoothly through the sonic transition. As a result, we consider 
the limiting case of an isothermal equation of state $n \to \infty$ 
for much of this paper. 

\subsection{Dimensionless Formulation} 

The problem can be re-formulated using dimensionless variables.
Length scales can be measured in terms of the stellar radius $R_\ast$.
We also define a reference scale $\rho_1$ for the density and use the
corresponding sound speed $a_1$ as a reference velocity, given by
$a_1^2 = [\partial P / \partial \rho]_1$, or, equivalently,
\be
a_1^2 \equiv K \left( 1 + {1 \over n} \right) \rho_1^{1/n} \, . 
\ee
We can then define the following dimensionless quantities 
\be
\alpha \equiv {\rho \over \rho_1} , \qquad 
u \equiv {|{\bf u}| \over a_1} , \qquad 
\qquad \xi \equiv {r \over R_\ast} , \qquad {\rm and} 
\qquad \psi \equiv {\Psi \over a_1^2} \, . 
\ee 
Note that $u$ is the (sonic) Mach number for isothermal flow.  The
stellar radius is typically $R_\ast \sim 10^{11}$ cm.  The reference
sound speed $a_1$ corresponds to that of the flow, which is launched
from the inner edge of the disk; the appropriate temperature is that
just above the disk surface, where the flow is exposed to UV heating,
so that $T\sim{10^4}$ K and $a_1\sim10$ km/s. The reference density
scale $\rho_1$ depends on the mass accretion rate ${\dot M}$; for the 
standard value ${\dot M} \approx 10^{-8} M_\odot$ yr$^{-1}$, and for 
typical values for the other quantities (see Section 5.3), the
reference density is given by $\rho_1/m_p = n_1 \sim 3 \times 10^{11}$
cm$^{-3}$.

Next we define a dimensionless parameter $b$ that measures the depth
of the gravitational potential well,
\be 
b \equiv {G M_\ast \over R_\ast a_1^2} \approx 667 
\left( {M_\ast \over 0.5 M_\odot} \right) 
\left( {R_\ast \over 10^{11} \, {\rm cm} } \right)^{-1} 
\left( {a_1 \over 10\,\,{\rm km}\,\,{\rm s}^{-1}} \right)^{-2} \, . 
\label{bdefine} 
\ee
The depth of the potential well is thus expected to lie in the range
$b\sim500-1000$. Finally, we define a dimensionless parameter
$\rotcon$ that measures the effects of rotation,
\be
\rotcon \equiv \left( {\Omega R_\ast \over a_1} \right)^2 = 
b \left( {\Omega \over \,\,\Omega_\ast} \right)^2\, , 
\label{rotdefine} 
\ee
where $\Omega$ is the rotation rate of the star and $\Omega_\ast^2$ 
= $GM_\ast/R_\ast^3$ is the Keplerian rotation rate due to stellar 
gravity evaluated at the stellar surface. As a result, the rotation
parameter can be written as $\rotcon=b\,(R_\ast/r_{\rm co})^3$, where
$r_{\rm co}$ is the co-rotation radius.  T Tauri stars exhibit a wide
range of rotation periods $P=2\pi/\Omega\sim2-12$ days \citep{herbst}.
For the median value $P\sim6$ days, the co-rotation radius falls at 
$\sim8 R_\ast$, so we consider the range $r_{\rm co}\sim6-10R_\ast$.  

For simplicity we generally assume that the co-rotation radius 
$r_{\rm co}$ is coincident with the inner disk edge $r_d$ (so that
$\rotcon=b/\xi_d^3$, where $\xi_d=r_d/R_\ast\sim6-10$). In practice,
accretion from the disk surface is launched from an annulus of finite
extent $\Delta{r}$, although the width is generally narrow
($\Delta{r}<r$; see Section 5.3). If $\Delta{r}\ll{r}$, this finite
width is not an issue.  If the annulus extends beyond the co-rotation
radius, the flow falls in the ``propeller'' regime where the star is
spun down (Ustyugova et al. 2006). On the other hand, if the annulus
extends inside the co-rotation radius, magnetic links to the regions
of the disk spinning faster than the star, as well as accretion of
disk material, act to spin-up the star, which will then increase its
rotation rate and move the co-rotation radius further inward. However,
induced changes in the location of the corotation radius, and/or the 
stellar spin, generally occur on long time scales. 

In terms of the dimensionless fields defined above, the 
continuity equation takes the form
\be
\alpha {\partial u \over \partial p} + 
u {\partial \alpha \over \partial p} = - 
{\alpha u \over h_q h_\phi} {\partial \over \partial p} 
\left( h_q h_\phi \right) \, , 
\label{derivcont} 
\ee
and the force equation becomes 
\be
u {\partial u \over \partial p} + {\alpha^{1/n} \over \alpha} 
{\partial \alpha \over \partial p} - \rotcon \xi \sin\theta 
|\nabla p|^{-1} \left( \xhat \cdot \phat \right)
= - {\partial \psi \over \partial p} 
= - {\partial \psi \over \partial \xi} {\partial \xi \over \partial p} 
= - {b \over \xi^2} {\partial \xi \over \partial p} \, . 
\label{derivforce} 
\ee
These equations can be integrated immediately to obtain the solutions 
\be
\alpha u h_q h_\phi = \lambda \, , 
\label{simplecont} 
\ee
and 
\be
{1 \over 2} u^2 + n \alpha^{1/n} + \psi = \varepsilon + 
\rotcon \int \xi \sin\theta \left( \xhat \cdot \phat \right) 
{dp \over |\nabla p|} \equiv \varepsilon + \rotcon I \, , 
\label{simpleforce} 
\ee
where the second equality defines the integral $I$ (this integral 
can be written in simple form; see {\bf Result 2} of Appendix 
\ref{sec:mathappendix}).  The quantity $h_q h_\phi$ in equation
(\ref{simplecont}) is proportional to the inverse of the magnetic
field strength (consistent with the third part of equation
[\ref{simple}]). The parameters $\lambda$ and $\varepsilon$ are
constant along streamlines, but are not, in general, the same for all
streamlines (they are functions of $q$).

\subsection{Transitions through the Critical Points} 

Critical points in the flow arise when the fluid speed is equal to the
transport speed. In general, magnetic media support three types of MHD
waves and hence allow for three types of critical points.  In the case
where the flow is confined to follow magnetic field lines, the problem
has only one possible critical point, which occurs where the flow
speed equals the sound speed.  In order for the flow to pass smoothly
through this sonic point, only particular values of the constant
$\lambda$ are allowed. The equations of motion (\ref{derivcont}) and
(\ref{derivforce}) provide two equations for the two unknowns
$\partial u/\partial p$ and $\partial \alpha/\partial p$; by solving
for these quantities and requiring that the functions are continuous
at the sonic point (e.g., Shu 1992), we find the required matching
conditions
\be
u^2 = \alpha^{1/n} \qquad {\rm and} \qquad 
{\alpha^{1/n} \over h_q h_\phi} {\partial \over \partial p} 
\left( h_q h_\phi \right) + \rotcon \xi \sin\theta 
|\nabla p|^{-1} \left( \xhat \cdot \phat \right) \, 
= {b \over \xi^2} {\partial \xi \over \partial p} \, . 
\ee
Note that these expressions remain valid in the isothermal limit 
where $n \to \infty$ and hence $\alpha^{1/n} \to 1$. 
We must thus evaluate the geometrical factor $\cal G$ defined by
\be
{\cal G} \equiv {1 \over h_q h_\phi} {\partial \over \partial p} 
\left( h_q h_\phi \right) = {1 \over h_p} 
{\partial h_p \over \partial p} \, , 
\label{gdef} 
\ee
where the second equality follows from {\bf Result 1} in Appendix
\ref{sec:mathappendix}.  Note that for radial flow in spherical
coordinates, where $p$ = $\xi$, this factor would have the familiar
form ${\cal G}=2/\xi$.

In general, the divergence operator, and the geometrical factor 
${\cal G}$ have units of $p^{-1}$, but the coordinate $p$ is not
necessarily a length scale. Nonetheless, we can define a dimensionless
parameter $\divnum$ that measures the degree of divergence through the
construction
\be
\divnum = \divnum (\xi,\theta) = \xi H \, 
\left( {\partial p \over \partial \xi} \right)^{-1} \, {\cal G} \, 
\quad \longrightarrow \quad 
\divnum (\xi) = {\xi \over h_p} {\partial h_p \over \partial \xi} \, , 
\label{divindex} 
\ee
where the function $H$ is defined by equation (\ref{bighdef}) in
Appendix \ref{sec:mathappendix}.  The intermediate expression is
general, whereas the final equality applies to the specific case where
the flow is confined to a field line so that the angle $\theta(\xi)$
is a known function of $\xi$.  With the definition (\ref{divindex}),
the index $\divnum$ is of order unity and is a slowly varying function
of the coordinates. Similarly, we define the function $\roterm$ that
specifies the rotational term in the force equation,
\be
\roterm = \xi \sin\theta \, 
\left( {\partial p \over \partial \xi} \right)^{-1} 
\left( \xhat \cdot \nabla p \right) \, . 
\label{termrotate} 
\ee
In terms of these functions, the matching condition at the sonic 
point takes the seemingly simple form 
\be
\alpha^{1/n} {\divnum \over \xi} 
+ \rotcon \roterm = {b \over \xi^2} \, . 
\label{matchsimple}
\ee
In general the quantities $\divnum$ and $\roterm$ are functions of
$(\xi,\theta)$. For flow along a magnetic field line, which are lines
of constant $q$, the angle $\theta$ is a specified function of the
radial coordinate, so that we obtain functions of a single variable
$\divnum (\xi)$ and $\roterm (\xi)$.  In the isothermal limit $n \to
\infty$, the density dependence drops out, so that all of the terms in
equation (\ref{matchsimple}) become known functions of the variable
$\xi$. For a general polytropic index, the two equations of motion
(evaluated at the sonic point), in conjunction with the two matching
conditions, allow one to solve for the four unknowns $\xi_s$,
$\alpha_s$, $u_s$, and $\lambda$.  After eliminating the other three
variables, the equation that specifies the sonic point has the form
\be 
{\left(b-\rotcon\roterm\xi^2\right) \over \xi\divnum} =
\left\{ 2 h_p^{-2} \left[ \left(n + {1 \over 2}\right) 
{\left(b-\rotcon\roterm\xi^2\right) \over \xi\divnum} - 
\rotcon I - {b\over\xi} + b - n \right] \right\}^{1/(2n+1)} \, .
\label{generalsonic} 
\ee
This equation applies to steady polytropic flow that follows any
magnetic field configuration, where the geometry is encapsulated in
the index $\divnum(\xi)$ of the divergence operator, the scale factor
$h_p$, the rotational term $\roterm(\xi)$, and its integral $I$. The
physical system is specified by the depth $b$ of the potential well,
the rotational speed (through $\rotcon$), and the polytropic index
$n$. Finally, the streamline of the flow is determined by the
intersection point with the equatorial plane, where this point (often
taken to be the inner disk edge $\xi_d$) specifies the value of the
coordinate $q$. Equation (\ref{generalsonic}) remains valid in the
isothermal limit $n \to \infty$, where the right hand side of the
equation becomes unity.

With the location of the sonic point $\xi_s$ specified by equation
(\ref{generalsonic}), the dimensionless mass accretion rate $\lambda$ 
for finite $n$ is given by 
\be
\lambda = h_p \left( 
{b - \rotcon \roterm \xi_s^2 \over \xi_s \divnum} 
\right)^{n + 1/2} \, , 
\ee
where all of the quantities are evaluated at $\xi_s$. In the
isothermal limit, accretion parameter $\lambda$ is specified 
implicitly through the transcendental equation
\be
\ln \lambda - {1 \over 2} \lambda^2 = \ln h_p + \rotcon I + 
b \left( {1 \over \xi_s} - 1 \right) - {1 \over 2} \, . 
\label{isolam} 
\ee

\subsection{General Constraint on Steady Polytropic Flow} 
\label{sec:genconstraint} 

In this section, we derive a general constraint that must be met for
steady transonic accretion to reach free-fall speeds in the inner
limit. Here, the flow starts at the inner disk edge (at subsonic
speed) and ends on the stellar surface (at supersonic speed). This
result applies to polytropic flow that follows magnetic field lines.
Let the integer $\ell$ denote the highest order multipole of the field
near the stellar surface and let $n$ denote the polytropic index of
the equation of state. We find that self-consistent transonic
solutions must satisfy the requirement
\be 
n > \ell + {3 \over 2} \, 
\label{flowconstraint} 
\ee
in order to achieve (nearly) free-fall speeds in the inner limit.
This constraint implies that, to leading order, the flow can be
described using an isothermal equation of state. 

The constraint of equation (\ref{flowconstraint}) can be derived as
follows: In the limit $\xi \ll 1$, only the highest order multipole
field contributes to the magnetic field. In this limit, the continuity
equation reduces to the form 
\be
\alpha u = \lambda A \xi^{-(\ell + 2)} \, , 
\ee
where $A$ is a constant (for example, this constant has values $A$ =
2, $2 \const$, and 2 for the magnetic field configurations considered
in Sections \ref{sec:dipole} and \ref{sec:dipoct}, and Appendix 
\ref{sec:diprad}, respectively). The force equation reduces to the form
\be
{1 \over 2} u^2 + n \alpha^{1/n} = {b \over \xi} \, . 
\ee
The combination of these two reduced equations of motion results in
the expression 
\be
{1 \over 2} u^2 + n \left( \lambda A \right)^{1/n} 
\xi^{-(\ell + 2)/n} u^{-1/n} = {b \over \xi} \, . 
\label{twoterms} 
\ee
In order for the solutions to approach a free-fall form in the inner
limit, the velocity field must approach the form $u^2 \sim 2b/\xi$,
which requires the first term to dominate the second in equation
(\ref{twoterms}).  Consistency thus requires
\be
u^2 \gg \alpha^{1/n} 
\qquad {\rm which} \, \, \, {\rm implies} \qquad 
{1 \over \xi} \gg {1 \over \xi^{(\ell + 3/2)/n}} \, . 
\ee
The requirement implies that $\ell + 3/2 < n$, as claimed in the
constraint of equation (\ref{flowconstraint}). 

We note that this argument applies for the limit where $\xi \to 0$. In
practice, magnetically controlled accretion takes place over a limited
range in radius, from the inner disk edge to the stellar surface,
i.e., spanning only about a factor of 10 in radial scale. In order for
the equation of state to be stiff enough to affect the accretion flow
over this more limited range, the polytropic index $n$ must be larger 
than the value $\ell + 3/2$, as given by equation (\ref{flowconstraint}).

In addition, for the case of dipole fields $\ell=1$, Koldoba et al.
(2002) find that the behavior of accretions flows from disks onto
magnetized stars has qualitatively different behavior when the
polytropic index $n$ is smaller or larger than $n=5/2=\ell+3/2$ (the
same threshold indicated by equation [\ref{flowconstraint}]). For
$n>5/2$, the Mach number increases monotonically as the flow
approaches the stellar surface (i.e., for decreasing radius); for
$n<5/2$, the Mach number first increases and then decreases as the
flow moves inward (see Koldoba et al. 2002 for further discussion).

The observational implications of this constraint are important: First
we note that current observations indicate that transonic flow with
free-fall speeds does take place.  Further, observations of magnetic
field signatures for T Tauri stars suggest that the octupole component
provides the largest contribution near the stellar surface.  As a
result, we should take $\ell$ = 3, so that the constraint becomes
$n>9/2$. From the inner disk edge to the stellar surface, the density
can vary significantly, e.g., by a factor of $\sim100$ for the models
considered in this paper.  Due to this constraint, however, the
temperature (sound speed) could vary by {\it at most} a factor of
$\sim2.8$ (1.7). As a result, the flow is close to isothermal, and we
consider the isothermal limit ($n \to \infty$) for most of this work.
This approximation is consistent with previous modeling work
\citep{har94}, which finds that reasonable hydrogen line profiles can
be obtained if the temperature profiles are roughly isothermal (except
for an initial rise in temperature where the flow leaves the disk).
On a related note, calculations of the thermal structure of accretion
funnels (Martin 1996) indicate that the Ca II and Mg II ions have a
strong cooling effect on the flow, i.e., they act as a thermostat and
give rise to nearly isothermal conditions in the vicinity of the
stellar surface.

\subsection{Mass Accretion Rate} 

The discussion thus far has focused on one streamline at a time.  
This section shows how the streamlines add up to determine the mass
accretion rate from the disk onto the star.  

The total mass accretion rate ${\dot M}_d$ leaving the inner portion
of the disk takes the form
\be
{\dot M}_d = 2 \int_{r_d}^{r_2} 2 \pi r dr \, \rho \, v \, = 
4 \pi R_\ast^2 \rho_1 a_1 \int_{\xi_d}^{\xi_2} \, \varpi d\varpi \, 
\alpha_1 u_1 = 4 \pi R_\ast^2 \rho_1 a_1 
\int_{\xi_d}^{\xi_2} \varpi d\varpi \lambda \, 
\left( h_p^{-1} \right)_d \, , 
\label{mdotdisk} 
\ee
where $r_d$ is the inner disk edge, the accreting annulus extends from
$r_d$ to $r_2$, and the variable $\varpi \equiv r/R_\ast$. The leading
factor of 2 arises because material accretes from both the top and
bottom surfaces of the disk.  In the final equality we have used the
boundary conditions that $\alpha$ = 1 and $u$ = $\lambda/h_p$ at the
start of the trajectory, and the subscript indicates that the quantity
is to be evaluated in the disk plane. The flow is assumed to leave the
disk surface in the vertical ($\zhat$) direction, consistent with the
magnetic field line geometries of interest (primarily dipoles and
octupoles).  For the sake of completeness, however, we note that the
flow is somewhat more complicated because the magnetic field threading
the disk bends away from the vertical at the disk surface, and because
the surface itself is not perfectly flat. As a result, material in the
accretion flow must first climb over a potential maximum before
leaving the disk and falling toward the star (Scharlemann 1978, Ghosh
\& Lamb 1979). In practice, of course, we see ample observational
evidence for accretion columns and magnetically truncated disks (see
Section \ref{sec:intro} for references), so that material can readily
leave the plane of the disk.

The streamlines, which follow the magnetic field lines, end on the
stellar surface, where the mass accretion rate ${\dot M}_\ast$ onto
the star is given by
\be
{\dot M}_\ast = 2 \int_{\mu_1}^{\mu_2} 2 \pi R_\ast^2 \, d\mu \, 
\rho \, v \, \left( \rhat \cdot \phat \right) = 
4 \pi R_\ast^2 \, \rho_1 a_1 \, \int_{\mu_1}^{\mu_2} d\mu \, 
\left( \rhat \cdot \phat \right)_\ast \, \lambda \, 
\left( h_p^{-1} \right)_\ast \, , 
\label{mdotstar} 
\ee
where the subscripts denote that the quantities are to be evaluated at
the stellar surface. The end points of integration $\mu_1$ and $\mu_2$
(where $\mu=\cos\theta$) are determined by the angles where the
streamlines (field lines) starting at $\xi_d$ and $\xi_2$ intersect
the stellar surface. The leading factor of 2 arises because accretion
onto the star takes place into rings on both hemispheres.  Notice that
the dot product $(\rhat\cdot\phat)$ must be included because, in
general, the streamlines do not intersect the stellar surface in the
radial direction.

In order for mass to be conserved, as required, the mass accretion
flow leaving the disk surface, given by equation (\ref{mdotdisk}),
must be the same as the mass accretion rate onto the stellar surface,
given by equation (\ref{mdotstar}). Comparison of the two expressions
shows that they are equal if the integrals are equal. This equality
holds and is shown in Appendix \ref{sec:conservemass} (which thus
provides a consistency check on this approach).

\section{Magnetic Accretion with a Dipole Field} 
\label{sec:dipole} 

This section considers the field geometry to be that of a simple
dipole. Although magnetic accretion with a dipole field has been
addressed previously (e.g., Hartmann et al. 1994, Koldoba et al.
2002), we revisit the problem using our approach, which sets up the
generalizations in the following section. 

\subsection{The Coordinate System} 

For the case of a pure dipole field, we use coordinates of the form 
\be
p = - \xi^{-2} \cos \theta \qquad {\rm and} \qquad 
q = \xi^{-1} \sin^2 \theta \, , 
\label{dipoledef} 
\ee 
where $\xi = r/R_\ast$ (see also Radoski 1967, Canalle et al. 2005,
and Appendix \ref{sec:mathappendix}).  The streamline (and hence
magnetic field line) that intersects the disk at its inner edge is
thus given by $q$ = $R_\ast/r_d$, so that this streamline corresponds
to the trajectory $\xi R_\ast$ = $r_d \sin^2 \theta$.

The covariant basis vectors $\epj$ arise from the gradients of the
scalar fields that define the coordinates.  If we express these
vectors in terms of the original spherical coordinates
$(\xi,\theta,\phi)$, the basis takes the form
\be
\epp = 2 \xi^{-3} \cos\theta \, \rhat + 
\xi^{-3} \sin\theta \, \thetat \, , 
\ee
\be
\epq = - \xi^{-2} \sin^2\theta \, \rhat + 
2 \xi^{-2} \sin\theta \cos\theta \, \thetat \, , 
\ee
and 
\be
\epphi = {1 \over \xi \sin\theta} \, \phihat \, , 
\label{phivector} 
\ee 
where the quantities $(\rhat, \thetat, \phihat)$ are the usual unit
vectors for spherical coordinates. The scale factors thus become
\be
h_p = \xi^{3} \left[ 4 \cos^2\theta + \sin^2\theta \right]^{-1/2} \, , 
\ee
\be 
h_q = {\xi^{2} \over \sin \theta} 
\left[ 4 \cos^2\theta + \sin^2\theta \right]^{-1/2} \, , 
\ee
and 
\be
h_\phi = \xi \sin \theta \, . 
\label{phifactor} 
\ee
The geometrical factor ${\cal G}$ in the divergence operator is thus 
\be
{\cal G} = 3 \xi^2 \cos \theta 
{ 8 - 5 \sin^2 \theta \over (4 - 3 \sin^2 \theta)^2} \, . 
\ee 
For flow along a field line (constant $q$), the effective index
$\divnum$ of the divergence operator (see equation [\ref{divindex}])
thus takes the form
\be
\divnum = \divnum (\xi) = {3 \over 2} \, 
\left( { 8 - 5 q \xi \over 4 - 3 q \xi} \right) \, . 
\ee
In the inner limit where $\xi \to 0$, the index $\divnum \to 3$, as
expected for dipole geometry. For streamlines, and hence magnetic
field lines, that connect the disk plane to the stellar surface, the
product $q \xi \to 1$ where the streamlines intersect the disk. As a
result, at the outer starting point of a streamline, the index
approaches the not-so-obvious value $\divnum \to 9/2$. Finally, for
dipole field configurations, the rotational term $\roterm$ (see
equation [\ref{termrotate}]) takes the simple form
\be
\roterm = \roterm (\xi) = {3 \over 2} q \xi^2 \, . 
\ee

\subsection{Isothermal Accretion Flow} 

We now consider isothermal accretion flow. For a given streamline
labeled by the coordinate $q$, we can eliminate the angular dependence
from the equation of motion. Even though the fluid fields are
functions of the coordinate $p$ only, the resulting expressions are
simpler in terms of the radial variable $\xi(p)$. This change of
variables is transparent as long as $\xi$ is monotonic as a function
of $p$. The equations of motion thus take the form
\be 
\alpha {\partial u \over \partial \xi} + 
u {\partial \alpha \over \partial \xi} =  
- \alpha \, u \, {\divnum \over \xi} = 
- \alpha \, u \, {3 \over 2 \xi} \, 
{8 - 5 q \xi \over 4 - 3 q \xi} \, , 
\ee
and 
\be 
u {\partial u \over \partial \xi} + {1 \over \alpha} 
{\partial \alpha \over \partial \xi} =  
- {b \over \xi^2} + \rotcon \, \roterm \, = 
- {b \over \xi^2} + {3 \over 2} \, \rotcon \, q \, \xi^2 \, . 
\ee
The matching condition for the sonic point takes the form 
\be
3 \xi \, {8 - 5 q \xi \over 4 - 3 q \xi} = 
2 b - 3 \, \rotcon \, q \, \xi^4 \, .
\label{dipsonic} 
\ee 
For the case where the inner disk edge corresponds to the co-rotation
point, the solution to equation (\ref{dipsonic}) lies just inside the
point $\xi_0 = (2/3)^{1/4}$ $r_d/R_\ast$.  We denote the sonic point
as $\xi_s$.

The equations of motion can be integrated to take the forms  
\be
\alpha u = \lambda (q \xi)^{-3} \left( 4 - 3 q \xi \right)^{1/2} \, , 
\label{dipcont} 
\ee
and 
\be
{1 \over 2} u^2 + \ln \alpha = \varepsilon + {b \over \xi} + 
{1 \over 2} \rotcon q \xi^3 \, . 
\label{dipforce} 
\ee
At the inner disk edge, $\xi=1/q$ and $\alpha=1$; the integration
constant $\lambda$ in equation (\ref{dipcont}) is defined so that
$u=\lambda$ at this boundary.  The second constant $\varepsilon$ 
from equation (\ref{dipforce}) is then given by 
\be
\varepsilon = {1 \over 2} \lambda^2 - {3 \over 2} b q \, , 
\ee
where we have assumed that the inner disk edge, the launching point of
the flow, and the co-rotation radius coincide (so that $\rotcon=bq^3$).
Using this result to specify $\varepsilon$ and taking the logarithm 
of the continuity equation (\ref{dipcont}), we find the following
implicit specification of the remaining constant $\lambda$: 
\be
\ln \lambda - {1 \over 2} \lambda^2 = 3 \ln \xi_s - {1 \over 2}
\ln (4 - 3 \xi_s) - {1 \over 2} + b \left( {1 \over \xi_s} + 
{1 \over 2} \xi_s^3 - {3 \over 2} \right) \, .  
\ee
This equation has two roots for $\lambda$, one with $\lambda<1$ and
another with $\lambda>1$. The accretion solutions of interest here
start with subsonic speeds at $\xi=\xi_d=1/q$, where $u=\lambda$, so
we must take the smaller root ($\lambda<1$).

\begin{figure} 
\figurenum{2} 
{\centerline{\epsscale{0.90} \plotone{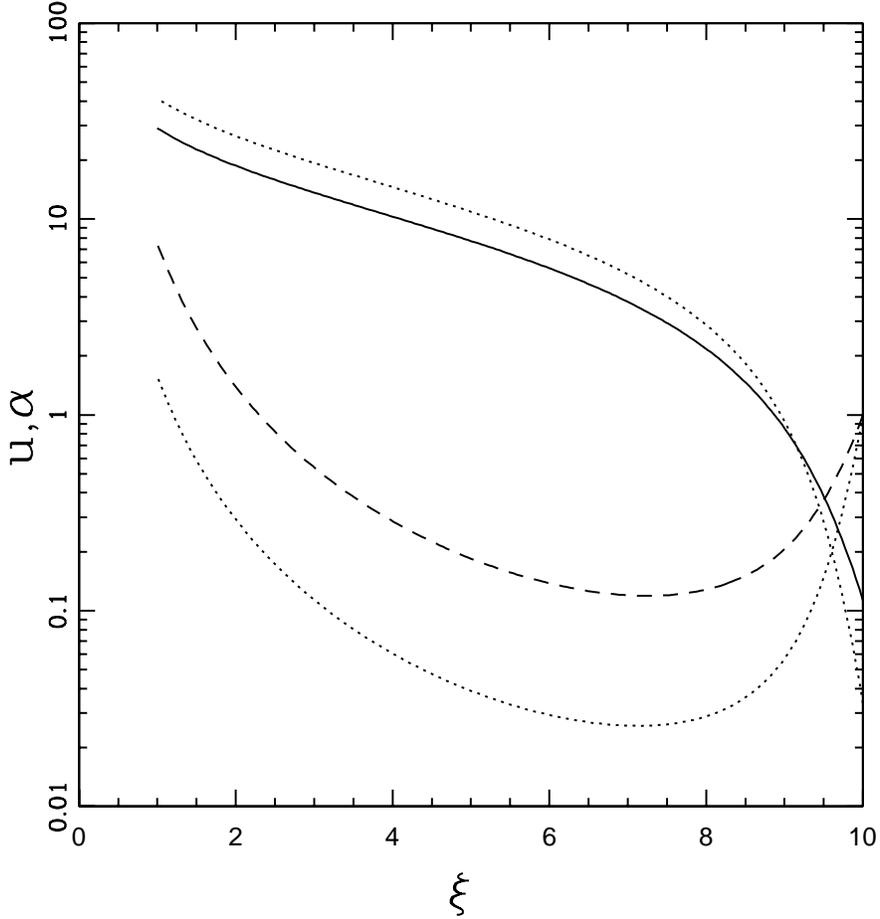} } } 
\figcaption{Dimensionless fluid fields for accretion solution using a
pure dipole geometry and an isothermal equation of state. The inner
disk edge lies at $\xi = \xi_d$ = 10 and the stellar surface
corresponds to $\xi = \xi_\ast$ = 1.0.  For dimensionless depth of the
gravitational potential well $b$ = 500, the velocity field $u$ is
shown by the solid curve and the density $\alpha$ is shown by the
dashed curve. The corresponding solutions for $b$ = 1000 are shown by
the dotted curves. For the reference values $n_1$ = $3 \times 10^{11}$
g cm$^{-3}$ and $a_1$ = 10 km s$^{-1}$, these solutions have physical 
speeds $u_\ast\approx$ 290 km s$^{-1}$ (410 km s$^{-1}$) and number
densities $n_\ast\approx 2\times10^{12}$ g cm$^{-3}$ ($5\times10^{11}$
g cm$^{-3}$) at the stellar surface for $b$ = 500 (1000). }
\label{fig:dipoleflow} 
\end{figure}

For flow that passes smoothly through the sonic point, we can find the
dimensionless fluid fields $\alpha$ and $u$.  These profiles are
plotted in Figure \ref{fig:dipoleflow} for two choices of the
dimensionless depth $b$ of the gravitational potential well of the
star.  Notice that the density field initially decreases inward from
the starting point (at the disk edge), but eventually increases.  This
behavior can be understood by finding the limiting form of the
solutions for the regimes $|1 - \xi| \ll 1$ and $\xi \ll 1$, as shown
in the following subsection.

\subsection{Limiting Forms for the Flow Solutions} 

In the regime $|1 - q\xi| \ll 1$, we define $\eta$ such that 
$q\xi = 1 - \eta$, where $\eta \ll 1$ (and $q=1/\xi_d$). 
The equations of motion take the form 
\be
\alpha u = \lambda \left( 1 + {9 \over 2} \eta \right) \, , 
\ee
and
\be
{1 \over 2} u^2 + \ln \alpha = {1 \over 2} 
\left( \lambda^2 - bq \eta \right) \, , 
\ee
where we have kept only the leading order terms (in $\eta$).
Working to leading order (in $\eta$), we find 
\be
u \approx \lambda \exp \left[ {1\over 2} \eta (bq + 9) \right] 
\qquad {\rm and} \qquad 
\alpha = \left( 1 + {9 \over 2} \eta \right) 
\exp \left[ - {1\over 2} \eta (bq + 9) \right] \, .
\ee 
These expressions can be expanded further, keeping only the leading
order terms in $\eta$, to obtain the forms 
\be
u \approx \lambda \left[ 1 + {1\over 2} \eta (bq + 9) \right]  
\qquad {\rm and} \qquad 
\alpha \approx 1 - {1 \over 2} bq \eta \, . 
\ee
The initial decrease in the density field is thus clear
(see Figure \ref{fig:dipoleflow}). 

In the opposite limit where $\xi \ll 1$, the equation of motion 
reduce to the forms 
\be
\alpha u = 2 \lambda q^{-3} \xi^{-3} \, , 
\ee
and 
\be
{1 \over 2} u^2 + \ln \alpha = {b \over \xi} \, . 
\ee
To leading order, the dimensionless fluid fields become
\be
u \approx \left( {2 b \over \xi} \right)^{1/2} 
\qquad {\rm and} \qquad 
\alpha \approx \left( {2 \over b} \right)^{1/2} 
{\lambda q^{-3} \over \xi^{5/2} } \, . 
\ee
The fluid fields thus increase as $\xi \to 0$. 

One can also show that the velocity field $u(\xi)$ is monotonic for
this problem. If we use the equations of motion to eliminate the
density $\alpha$ in favor of the velocity $u$, we obtain 
\be
{1 \over 2} u^2 - \ln u = \varepsilon + {b \over \xi} + 
{1 \over 2} \, b \, q^4 \, \xi^3 - \ln \lambda + 3 \ln (q\xi) 
- {1 \over2} \ln \left[ 4 - 3 q \xi \right] \, .  
\ee
Taking the derivative of both sides we find 
\be
\left( u - {1 \over u} \right) {du \over d \xi} = 
- {b \over \xi^2} + {3 \over 2} \, b \, q^4 \, \xi^2 + 
{3 \over \xi} + {3 q \over 2(4 - 3 q \xi)} \equiv F(\xi) \, .  
\label{uslope} 
\ee
One can show that the function $F(\xi)$, the right hand side 
of the above expression, is monotonically increasing on the 
interval $0 \le \xi \le 1$, has a zero at the sonic point, 
is positive for $\xi = 1$ and is negative in the limit 
$\xi \to 0$. These properties, in conjunction with equation 
(\ref{uslope}) imply that $u$ is monotonic and decreasing. 
Note that the statement that $u$ is decreasing means that 
the velocity increases as fluid elements approach the star. 

\section{Magnetic Accretion with Dipole plus Octupole Field} 
\label{sec:dipoct} 

In this section we consider the stellar magnetic field to have both
dipole and octupole components. The magnetic field thus takes the form
\be
{\bf B} = {B_{\rm oct} \over 2} \xi^{-5}
\left[ \left(5 \cos^2\theta - 3 \right) \cos\theta \, \rhat 
+ {3 \over 4} \left(5 \cos^2\theta - 1 \right) \sin\theta \, \thetat 
\right] + {B_{\rm dip} \over 2} \xi^{-3} 
\left( 2 \cos \theta \, \rhat + \sin\theta \, \thetat \right) \, , 
\label{dipoctfield} 
\ee 
where $\xi=r/R_\ast$ is the dimensionless radius. The leading factors
of 1/2 (for both the dipole and octupole terms) are included to be
consistent with the convention of Gregory et al. (2010).  If we scale
out the dipole field strength, the relative size of the octupole
contribution is given by the dimensionless parameter
\be 
\const \equiv {B_{\rm oct} \over B_{\rm dip}} \, .  
\ee 
Observations of the signatures of magnetic accretion onto T Tauri
stars indicate that the parameter $\const$ lies in the range
$0\le\const\le10$. For example, the young stars V2129 Oph (Donati et
al. 2007) and BP Tau (Donati et al. 2008) are observed to have field
parameter $\const=1-4$. The star AA Tau (Donati et al. 2010b) has a
nearly dipole field with $\const\approx0.25$, whereas TW Hya has a
much larger octupole component with $\const\approx4$ observed at one
epoch and $\const\approx6$ found at another (Donati et al. 2011b).

\subsection{The Coordinate System} 

With the configuration of equation (\ref{dipoctfield}), the magnetic
field is current-free and curl-free, and can be written as the
gradient of a scalar field (see Appendix \ref{sec:mathappendix}).  
The first scalar field $p$ of the coordinate system takes the form
\be
p = - {1 \over 4} \xi^{-4} \const \left( 5 \cos^2\theta - 3 \right) 
\cos\theta - \xi^{-2} \cos\theta \, . 
\label{pdef} 
\ee
The gradient $\nabla p$ points in the direction of the magnetic
field. Next we construct the perpendicular vector field $\nabla q$,
where the scalar field $q$ provides the second coordinate and 
is given by 
\be
q = {1 \over 4} \xi^{-3} \const \left( 5 \cos^2\theta - 1 \right) 
\sin^2\theta + \xi^{-1} \sin^2\theta \, . 
\label{qdef} 
\ee 
The scalar fields $(p,q)$ represent orthogonal coordinates in the
poloidal plane and are used instead of the spherical coordinates
$(\xi,\theta)$.  In this version of the problem, the magnetic field is
axisymmetric about the $\zhat$ axis, so that the usual azimuthal
coordinate $\phi$ is the third scalar field. Note that both $p$ and
$q$ are dimensionless.

Next we find the covariant basis vectors $\epj$, which can be written
in terms of the original coordinates $(\xi,\theta)$, so that the 
basis takes the form
\be
\epp = \left[ \xi^{-5} \const \left( 5 \cos^2\theta - 3 \right) + 
2 \xi^{-3} \right] \cos\theta \, \rhat + 
\left[ {3 \over 4} \xi^{-5} \const \left( 5 \cos^2\theta - 1 \right) +  
\xi^{-3} \right] \sin\theta \, \thetat \, , 
\ee
and 
\be
\epq = 
- \left[ {3 \over 4} \xi^{-4} \const \left( 5 \cos^2\theta - 1 \right) + 
\xi^{-2} \right] \sin^2\theta \, \rhat + 
\left[ \xi^{-4} \const \left( 5 \cos^2\theta - 3 \right) + 
2 \xi^{-2} \right] \sin\theta \cos\theta \, \thetat \, , 
\ee
where the third basis vector $\epphi$ is given by equation
(\ref{phivector}). 

It is useful to define ancillary functions 
\be
f = \const \left( 5 \cos^2\theta - 3 \right) + 2 \xi^2 
\qquad {\rm and} \qquad 
g = {3 \over 4} \const \left( 5 \cos^2\theta - 1 \right) 
+ \xi^2 \, . 
\ee
With these definitions, we can write the magnitudes of the basis
vectors in the forms
\be
|\epp|^2 = \xi^{-10} \left[ f^2 \cos^2\theta + g^2 
\sin^2\theta \right] \, , 
\ee
and 
\be
|\epq|^2 = \xi^{-8} \sin^2 \theta 
\left[ g^2 \sin^2\theta + f^2 \cos^2\theta \right] \, . 
\ee
The corresponding scale factors thus become 
\be
h_p = \xi^{5} \left[ f^2 \cos^2\theta + g^2 
\sin^2\theta \right]^{-1/2} \, , 
\ee
and 
\be 
h_q = {\xi^{4} \over \sin \theta} 
\left[ g^2 \sin^2\theta + f^2 \cos^2\theta \right]^{-1/2} \, , 
\ee
where the third scale factor $h_\phi$ is given by equation
(\ref{phifactor}).

This specification of the divergence operator is implicit. One could
invert equations (\ref{pdef}) and (\ref{qdef}), and then write the
spherical coordinates $(\xi,\theta)$, the scale factors $(h_p, h_q,
h_\phi)$, and the ancillary functions $(f,g)$ as functions of the new
coordinates $(p,q)$. However, the definitions of $(p,q)$ are
nontrivial functions of $(\xi,\theta)$, so that inversion is
complicated and unwieldy.  For clarity, we leave this construction in
implicit form.

\begin{figure} 
\figurenum{3} 
{\centerline{\epsscale{0.90} \plotone{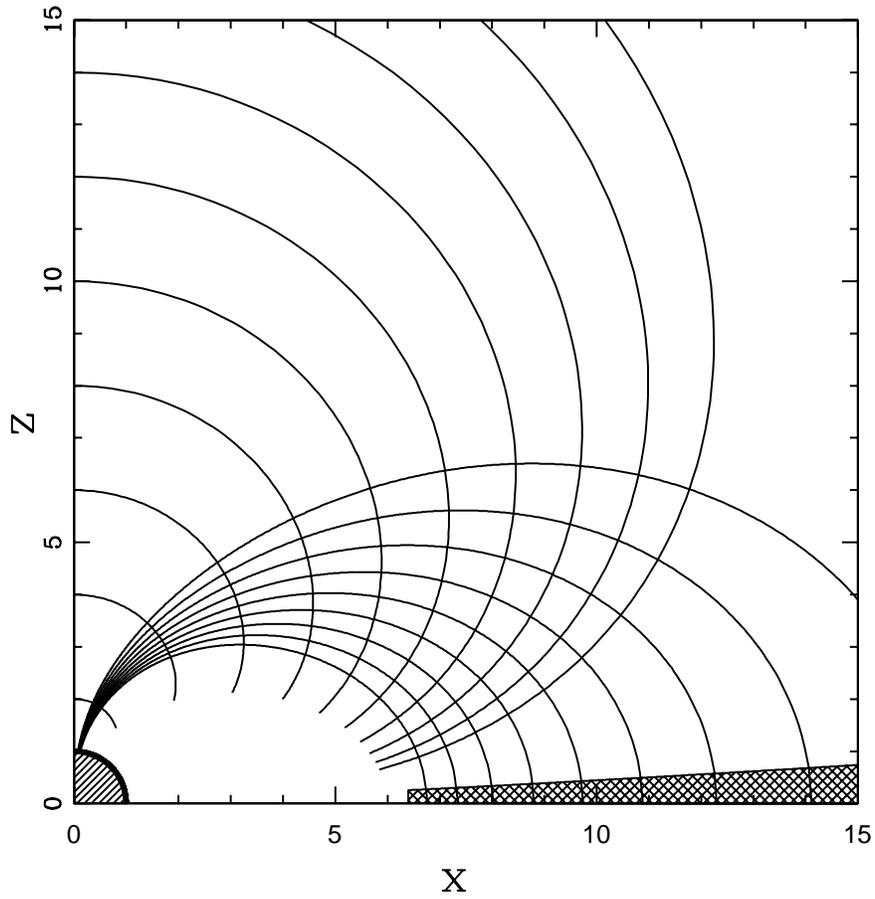} } } 
\figcaption{Magnetic field lines centered on the star for 
configurations with both a dipole and an octupole contribution. 
For this case, the dimensionless parameter $\const$ = 10. The star is
shown in the lower left of the figure, the disk is shown as a wedge
just above the $x$-axis. A collection of lines of constant $p$, the
surface orthogonal to the field lines, is also shown. }
\label{fig:blines} 
\end{figure}

\subsection{Results for Flow Along a Field Line} 

The lines of constant $q$ define the trajectories of fluid elements
following along magnetic field lines. Here we are interested in the
subset of field lines that intersect the equatorial plane and hence
intersect the disk. Note that in a fixed poloidal plane, and close to
the stellar surface, the axisymmetric dipole plus octupole fields have
three regions of closed field line loops between the north and south
pole of the star; the higher latitude regions that do not intersect
the disk are not considered here. By inverting equation (\ref{qdef}),
we can find the angle $\theta$ as a function of radius $\xi$ along a
field line:
\be
\sin^2 \theta = {2 \over 5 \const} \left\{ (\xi^2 + \const) -
\left[ (\xi^2 + \const)^2 - 5 \const q \xi^3 \right]^{1/2} 
\right\} \, . 
\label{sinefunxi} 
\ee
For the streamlines of interest, we must choose the negative sign for
the discriminant (as shown below in equation [\ref{qtruncate}], for
sufficiently large values of $\const$, streamlines that intersect the
disk have $q < 0$, and we must use the opposite sign). The ancillary
functions can then be written as a function of the radial coordinate
only:
\be 
f (\xi) = 2 \left[ (\xi^2 + \const)^2 - 5 \const q \xi^3 \right]^{1/2} 
\label{fxi} 
\ee
and
\be
g (\xi) = {3 \over 2} \left\{ \const +
\left[ (\xi^2 + \const)^2 - 5 \const q \xi^3 \right]^{1/2} 
\right\}  - {1 \over 2} \xi^2 \, . 
\label{gxi} 
\ee
With these definitions, we can find the rotational function $\roterm$
(see equation [\ref{termrotate}]) for flow along a field line,
\be
\roterm (\xi) = {2 \xi \over 5 \const} \left\{ (\xi^2 + \const) - 
\left[ (\xi^2 + \const)^2 - 5 \const q \xi^3 \right]^{1/2} \right\}
\left\{ 1 + {g(\xi) \over f(\xi)} \right\} \, .
\label{rotermdioct}  
\ee
The index $\divnum$ (from equation [\ref{divindex}]) takes the form 
\be
\divnum (\xi) = 5 - \left[ f^2 + (g^2 - f^2) {1 \over 5 \const} 
\left( 2\xi^2 + 2 \const - f \right) \right]^{-1} \times 
\qquad \qquad \qquad \qquad \qquad \qquad 
\label{divindexdioct} 
\ee
$$
\, \qquad 
{\xi \over 5 \const} 
\left\{ 5 \const f f_\xi + 
\left[ g \left( {3 \over 4} f_\xi - \xi \right) - f f_\xi \right] 
\left( 2\xi^2 + 2 \const - f \right) + (g^2 - f^2) 
\left( 2 \xi - {1 \over 2} f_\xi \right) \right\} \, , 
$$
where $f(\xi)$ and $g(\xi)$ are given by equations (\ref{fxi},
\ref{gxi}), and where $f_\xi = df/d\xi$.

\begin{figure} 
\figurenum{4} 
{\centerline{\epsscale{0.90} \plotone{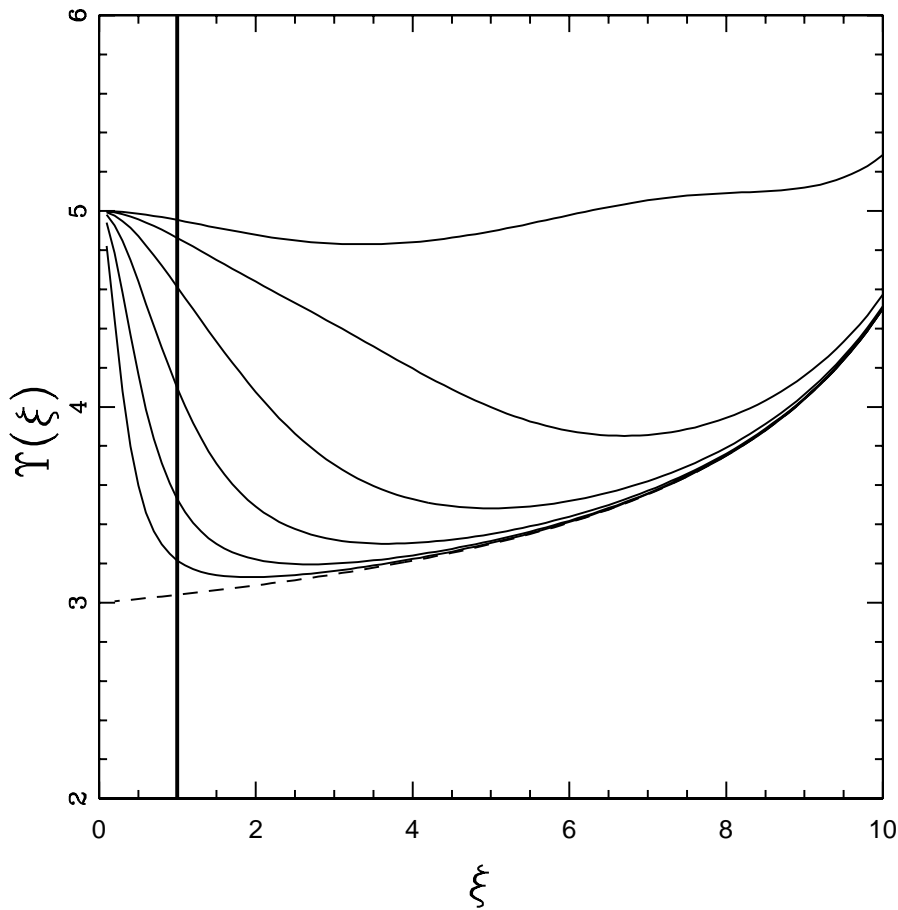} } } 
\figcaption{Index $\divnum$ of the divergence operator (see equations 
[\ref{divindex}] and [\ref{divindexdioct}]) for magnetic field
configurations with both dipole and octupole components. The relative
strength of the octupole component $\const$ has values given by
$\log_{10} \const$ = 3/2, 1, 1/2, 0, --1/2, and --1 (from top to
bottom). The dashed curve shows the limiting case of a pure dipole
field where $\const \to 0$. The heavy solid line at $\xi$ = 1 marks
the surface of the star, where accretion flow stops; the curves
continue inward to smaller values of $\xi$ to illustrate the
asymptotic behavior of $\divnum (\xi)$. }
\label{fig:index} 
\end{figure}

The index $\divnum$ is a slowly varying function of the radius $\xi$,
or, equivalently, the position $p(\xi)$ along the field line. The
index function $\divnum$ is plotted in Figure \ref{fig:index} for
varying values of the parameter $\const$ that sets the relative
strength of the octupole component.  The solid curves show the index
$\divnum (\xi)$ for parameter values $\const$ from $10^{-2}$ (bottom)
to $10^{3/2}$ (top). The dashed curve shows the limiting case of a pure
dipole field.  In the limit $\xi \to 0$, the index $\divnum \to 5$, as
expected for an octupole field. For a pure dipole field (dashed
curve), the limiting value $\divnum \to 3$.  Near the inner disk edge,
the index is somewhat larger than the value $\divnum$ = 9/2 expected
for a dipole. For the intermediate regime in $\xi$, the index first
decreases as $\xi$ decreases (toward the dipole value of 3), but the
index increases closer to the star as the octupole contribution
dominates. For sufficiently large values of $\const \gg 1$, the
octupole component dominates the field even near the inner disk edge;
for this regime, the value of the index approaches $\divnum$ = 15/2
(from equation [\ref{divindexdioct}] in the limit $\const \to
\infty$).  Notice also that for the value $\const \sim 30$ (e.g., the
top curve in Figure \ref{fig:index}), the compromise between the
dipole and octupole terms leads to the index being nearly constant 
with value $\divnum \approx 5$. 

The field lines cross the equatorial plane where $\sin\theta = 1$ and
$\cos\theta = 0$ (where the fluid trajectories start).  The value of
$q$ that leads to plane-crossing at radius $\xi_d$ is thus given by
\be
q = {1 \over \xi_d} 
\left( 1 - {\const \over 4 \xi_d^2} \right) \, . 
\label{qtruncate} 
\ee
The dimensionless truncation radius $\xi_d\sim10$, so we expect
$q\sim1/10$ for field lines that thread the disk.  If the octupole
contribution is too large, $\const > 4 \xi_d^2 \sim 100$, the field
lines that intersect the disk would have $q < 0$ and the sign of the
discriminant in equation (\ref{sinefunxi}) must change. T Tauri
systems are expected to have $\const\lta10$, so this complication
should not arise. 

Using the result $q \ll 1$, we can find an approximate expression for
the co-latitude on the star where fields lines starting at the disk 
truncation point reach the stellar surface, i.e., 
\be
\sin^2 \theta_\ast = {q \over 1 + \const} + {\cal O} (q^2) 
= {1 \over \xi_d (1 + \const) } + {\cal O} \left( \xi_d^{-2} \right) \, . 
\ee
The second equality uses equation (\ref{qtruncate}) to write $q$ in
terms of the disk truncation radius $\xi_d$.  This result shows that
the field lines that connect the disk truncation to the star must meet
the stellar surface near the pole $(\theta_\ast\ll{1})$. Increasing
the octupole contribution (via increasing $\const$) decreases the
angle $\theta_\ast$ even further.

We can also find the ratio of areas. Let ${\cal A}_d$ be the area of
an annulus on the disk. In the limit of a thin annulus with width
$\Delta \xi$, the area ${\cal A}_d$ = $2 \pi \xi (\Delta \xi)$. This
area gets funneled onto the stellar surface over a much smaller area
${\cal A}_\ast$ given by 
\be
{\cal A}_\ast = 2 \pi \left[ \cos \theta_{\ast 2} - 
\cos \theta_{\ast 1} \right] \, . 
\ee
This expression does not include the dot product $\rhat\cdot\phat$,
which takes into account the non-radial direction of the incoming
material (equation [\ref{mdotstar}] and Appendix \ref{sec:conservemass}). 
Since the angle $\theta_\ast\ll{1}$, the dot product is close to unity, 
$\rhat\cdot\phat\approx1-(3\const+1)^2/(8\xi_d[\const+1]^3)+\dots$,
and can be neglected to leading order.  Using the above results, we
can evaluate the ratio of areas to find
\be
{ {\cal A}_\ast \over {\cal A}_d } = 
{1 \over 2 \, (1 + \const) \, \xi_d^3 } \left\{ 1 + 
{\cal O} \left( \xi_d^{-2} \right) + 
{\cal O} \left( [\Delta \xi]^{2} \right) \right\} \, \, . 
\ee
This leading order expression shows that the ratio of areas decreases
inversely with the increase in field strength, which increases as
$\xi^{-3}$ from the dipole contribution and includes an extra factor
of $(1+\const)$ at the stellar surface due to the octupole component. 

In the limit of small $\xi$, we can find asymptotic forms for the
velocity and density fields. As long as the equation of state is not
too stiff, the dimensionless fluid fields approach the form
\be
u \sim \left( {2 b \over \xi} \right)^{1/2} 
\qquad {\rm and} \qquad \alpha \sim \left( {2 \over b} \right)^{1/2}
\lambda \, (\const + 1) \, \xi^{-9/2} \, . 
\label{inneroct}  
\ee
These forms are valid when the polytropic index $n > 9/2$ (see Section 
\ref{sec:genconstraint}). For flow with $n < 9/2$, free-fall
velocities are not realized.

\begin{figure} 
\figurenum{5} 
{\centerline{\epsscale{0.90} \plotone{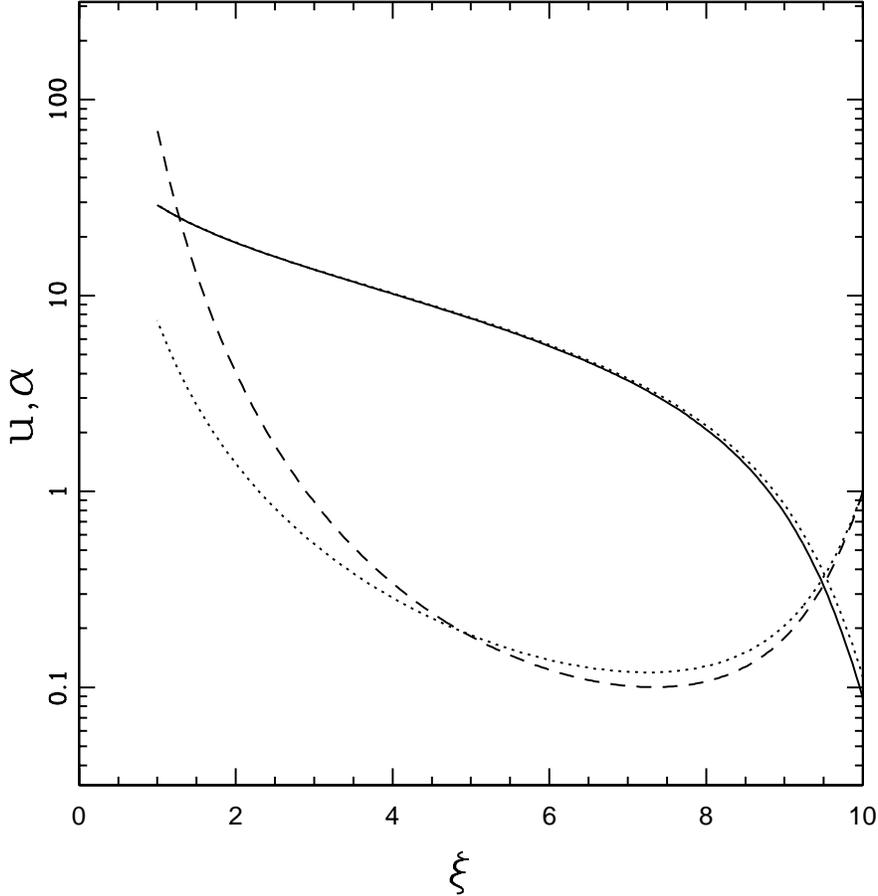} } } 
\figcaption{Dimensionless fluid fields for the velocity $u$ (solid 
curve) and density $\alpha$ (dashed curve) for accretion solution with
both dipole and octupole components.  In these units, the inner disk
edge corresponds to $\xi$ = $\xi_d$ = 10 and the stellar surface lies
at $\xi = \xi_\ast$ = 1.  The relative strength of the octupole
component is set by the parameter $\const$ = 10, and the dimensionless
depth of the gravitational potential well $b$ = 500.  The dotted
curves show the corresponding solution for a dipole field (in the
limit $\const \to 0$). For the reference values $n_1$ =
$3\times10^{11}$ g cm$^{-3}$ and $a_1$ = 10 km s$^{-1}$, these
solutions have speeds $u_\ast\approx$ 290 km s$^{-1}$ at the stellar
surface; the corresponding number densities are $n_\ast\approx
2\times10^{12}$ g cm$^{-3}$ for the pure dipole and $n_\ast\approx
2\times10^{13}$ g cm$^{-3}$ for the dipole/octupole configuration. } 
\label{fig:octfields} 
\end{figure}

\subsection{Accretion Solutions} 

Figure \ref{fig:octfields} shows the density and velocity fields for a
typical system with magnetic octupole parameter $\const$ = 10 and
dimensionless depth of the gravitational potential $b$ = 500.  The
solid curve shows the dimensionless velocity field $u(\xi)$ and the
dashed curve shows the corresponding density profile $\alpha(\xi)$.
For comparison, the dotted curves show the profiles for the same
system with a pure dipole field.  Since the dipole contribution
dominates the magnetic field at the inner disk edge, and since the
sonic point falls relatively near the edge, the location of the sonic
point $\xi_s$ and the mass accretion constant $\lambda$ are only
weakly dependent on the parameter $\const$. Specifically, the sonic
point $\xi_s \approx$ 8.80 for $\const$ = 10, compared with $\xi_s
\approx$ 8.87 for $\const \to 0$; similarly, the parameter $\lambda
\approx$ 0.0877 for $\const$ = 10, compared with $\lambda \approx$
0.113 for the limit $\const \to 0$.  Notice, however, this statement
no longer holds if the octupole component dominates the magnetic field
at the inner disk edge. 

The most important difference between the two field configurations
(with solutions shown in Figure \ref{fig:octfields}) is that the
density increases near the stellar surface more rapidly in the
presence of an octupole component; here the density is larger than
that of the dipole limit by a factor of $\sim 9.5$. Equation
(\ref{inneroct}) suggests that the density should be larger than that
of the dipole case by a factor of $(1+\const)=11$; this factor is
somewhat larger than the value (9.5) depicted in Figure
\ref{fig:octfields} because the solutions are not fully in the
$\xi\to0$ limit. Nonetheless, the inclusion of octupole field
components allows the flow densities at the stellar surface to be
greater by an order of magnitude compared to those from the dipole
limit.

The physical value of the flow density at the stellar surface, before
the accretion shock, can be estimated as follows. The mass accretion
rate can be written in the approximate form
\be
{\dot M} = 2 \pi \rho_1 a_1 (r_2 + r_d) (r_2 - r_d) \lambar \, , 
\ee 
where $\lambar$ is normalized so that $u=\lambda$ at the inner disk
edge and the mean value is taken to account for possible variations
over the range of launching radii on the disk surface. It is useful to
define a dimensionless width of the annulus where the flow originates,
i.e., $w\equiv(r_2-r_d)/r_d$. The number density $n_1$ at the start of
the flow can then be written in the form
\be
n_1 = {3\times10^{11}\,{\rm g}\,{\rm cm}^{-3} \over (1 + w/2) w} 
\left( { {\dot M} \over 10^{-8}\,M_\odot\,{\rm yr}^{-1}} \right) 
\left( {a_1 \over 10 \,{\rm km}\,{\rm s}^{-1}} \right)^{-1}
\left( {\lambar \over 0.1} \right)^{-1}
\left( {r_d \over 10^{12}\,{\rm cm} } \right)^{-2} \, .
\label{numdense} 
\ee
For dipole field configurations (see Figure \ref{fig:dipoleflow}), 
the number density $n_\ast$ at the stellar surface is larger than 
the initial density $n_1$ by a factor of $1-10$, so that
$n_\ast\approx10^{12}$ g cm$^{-3}$ (where we have taken
$(1+w/2)w\sim1$). For field configurations with octupole
contributions, the number density is larger by another factor of
$\sim(1+\const)$ (see Figure \ref{fig:octfields}), so we expect
$n_\ast\approx10^{13}$ g cm$^{-3}$ for systems with substantial
octupole components. The densities could be even larger if the initial
annulus on the disk is narrow ($w\ll1$).

For comparison, coronal densities for T Tauri stars are estimated to
be of order $n_{cor}\sim10^9-10^{11}$ g cm$^{-3}$ \citep{ness04,kast04}, 
and these values are considered too low to produce the observed soft
X-ray exess emission from these sources \citep{brick}. The X-ray
emission can arise from shock heated plasma with temperatures
$T\sim{3}\times10^6$ K and densities $n\sim10^{11}-10^{13}$ g cm$^{-3}$ 
\citep{arg11}. These required densities are thus comparable to those
expected from accretion flow (see above); more significantly, octupole
contributions $\const\ne0$ may be necessary to explain the upper end
of this range.

The expected accretion hot spot temperature $T_S$ can be written in
terms of the other system parameters such that
\be
T_S^4 = {G M_\ast {\dot M} \over R_\ast \sigma_B {\cal A}_\ast} 
\left( 1 - {R_\ast \over r_d} \right) = 
{2\over\pi} \,\, {\xi_d^3(1+\const)\over{w(2+w)}} \,\, 
{G M_\ast {\dot M} \over \sigma_B R_\ast r_d^2}  
\left( 1 - {1 \over \xi_d} \right) \, , 
\ee
where $\sigma_B$ is the Stefan-Boltzmann constant, and the other
quantities have been defined previously. Note that $T_S$ is the hot
spot temperature at the end of the accretion flow where material
arrives at the star and generates an optically thick shock. For the
same physical values used to evaluate equation (\ref{numdense}), we
obtain $T_S\approx7000$ K $(1+\const)^{1/4}$. Even for a strong
octupole component, $\const\sim10$, the hot spot temperature is only
$T_S\sim12,700$ K.  As a result, the hot spot temperatures for
accretion flow along non-dipolar magnetic field lines remain
consistent with the constraints established by Muzerolle et al. (2001)
from line profile modeling, line ratios, and continuum emission
constraints (see their Figure 16 and associated discussion). We note
that the higher temperatures derived form X-ray observations of
He-like line triplets (e.g., Argiroffi et al. 2011), and discussed in
the previous paragraph, refer to the hotter post-shock regions.
Although a full treatment of X-ray signatures is beyond the scope of
this present work, other studies have suggested that emitted X-rays
from the accretion shock could be absorbed within the accretion funnel
(Sacco et al. 2010).

\begin{figure} 
\figurenum{6} 
{\centerline{\epsscale{0.90} \plotone{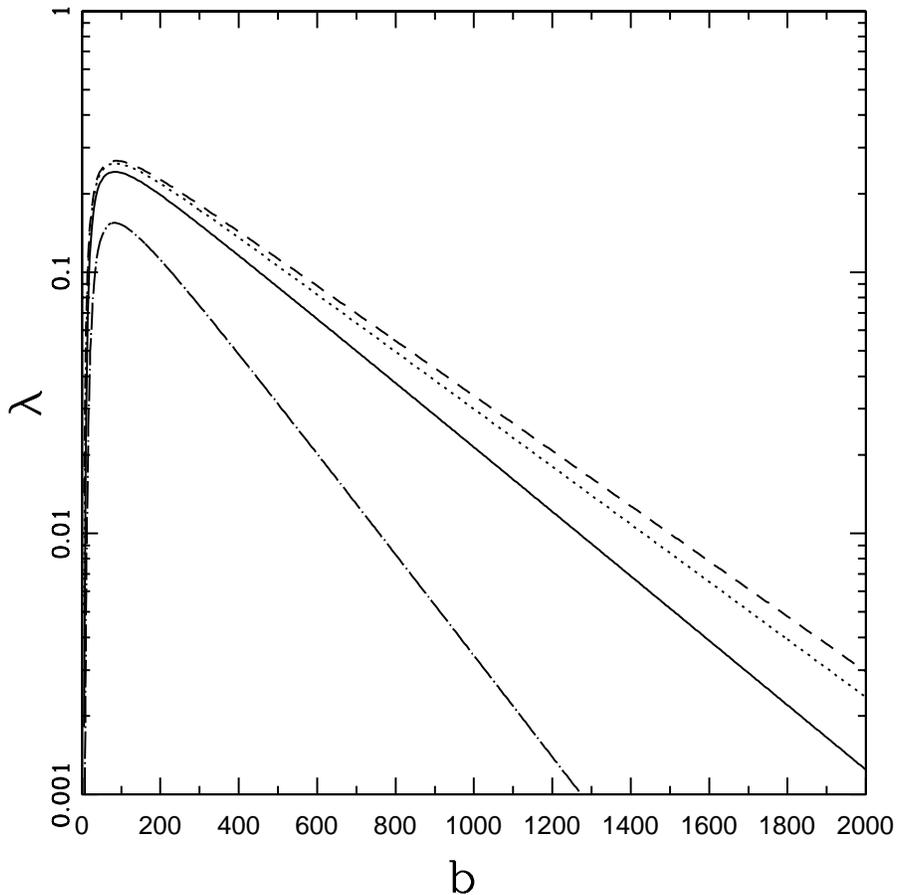} } } 
\figcaption{Dimensionless mass accretion parameter $\lambda$ as a
function of the dimensionless depth $b$ of the gravitational potential
well of the star. The magnetic field has both dipole and octupole
contributions, where the parameter $\const$ specifies the relative
strength of the octupole component.  The curves show $\lambda(b)$ for
varying values $\const$ = 30 (dot-dashed), 10 (solid), 3 (dotted), and
in the pure dipole limit $\const \to 0$ (dashed). } 
\label{fig:octlambda} 
\end{figure}

Figure \ref{fig:octlambda} shows the mass accretion parameter
$\lambda$ as a function of the depth $b$ of the stellar gravitational
potential well.  Curves are shown for a range of values for the
parameter $\const$ that sets the relative strength of the octupole
component of the magnetic field. As shown in the figure, increasing
the strength of the octupole contribution results in a decrease in the
mass accretion $\lambda$. For all values of $\const$, the parameter
$\lambda$ decreases in both the limit of large $b$ and the limit of
small $b$.  For sufficiently large $b$, Figure \ref{fig:octlambda}
shows that the accretion parameter $\lambda$ is an exponentially
decreasing function of the gravitational potential $b$. This behavior
can be understood as follows: In this regime, the matching condition
for the sonic point from equation (\ref{generalsonic}) reduces to the
form $\rotcon\roterm\xi^2=b$; since $\rotcon\propto{b}$, the location
of the sonic point $\xi_s$ becomes independent of the potential $b$.
The mass accretion parameter is specified through equation
(\ref{isolam}), which takes the form $\ln\lambda=A-Cb$, where $A$ and
$C$ are constant and $C=1-1/\xi_s>0$.  The mass accretion parameter is
then given by $\lambda\sim\exp[-Cb]$, as depicted in Figure
\ref{fig:octlambda}.

\section{Magnetic Truncation Radius}
\label{sec:truncate} 

In these accreting systems, the circumstellar disk is (indirectly)
observed to have an inner boundary at radius $r_d>R_\ast$. Further,
the inner disk can be truncated at radius $r_T$ where the stress due
to the stellar magnetic field is large enough to remove angular
momentum from the Keplerian flow (for further discussion, see Ghosh \&
Lamb 1979, K{\"o}nigl 1991, Shu et al. 1994). Here we assume that
$r_d=r_T$, but more complicated possibilities remain. The requirement 
of disk truncation implies a constraint on the magnetic field of the
form
\be 
B^2 \left( r, \theta = {\pi \over 2} \right) \approx 
4 \pi \rho v^2 \equiv \alpha { {\dot M} \over r^2} 
\left( {G M_\ast \over r} \right) ^{1/2} \, , 
\label{truncondition} 
\ee
where $\alpha$ is a dimensionless parameter of order unity; for
example, recent numerical simulations (Long et al. 2005) suggest that
$\alpha$ = 1/2 (see also the review of Bouvier et al. 2007, and
references therein). This parameter incorporates the difference
between exact equality in the two sides, the difference between the
flow speed and the free-fall speed, and the departure of the geometry
from spherical symmetry.

For a dipole field, we can write $B(r)$ = (1/2) $B_{\rm dip}$
$(R_\ast/r)^3$ for the magnetic field strength at the equatorial
plane. Equation (\ref{truncondition}) then reduces to the form
\be
r_T = \alpha^{-2/7} { \left( B_{\rm dip} R_\ast^3 /2 \right)^{4/7} 
\over \left( G M_\ast {\dot M}^2 \right)^{1/7} } \, .  
\label{trundipole} 
\ee
For typical parameters of T Tauri stars, the truncation radius 
$\xi_T = r_T / R_\ast$ = 5 -- 10. 

For a magnetic field with both dipole and octupole components, 
the truncation radius is given by the solution to the equation 
\be
\xi^{7/2} \left[ 1 - {3 \over 4} \const \xi^{-2} \right]^{-2} = 
\alpha^{-1} { B_{\rm dip}^2 R_\ast^2 \over 4 {\dot M} } 
\left( {R_\ast \over G M_\ast} \right) ^{1/2} = \xi_{T0}^{7/2} \, , 
\label{trundipoct} 
\ee
where the quantity $\xi_{T0}$ is the dimensionless truncation radius
for a pure dipole field. For the standard (aligned) field orientation
with $\const > 0$, equation (\ref{trundipoct}) shows that the magnetic
truncation radius is {\it smaller} for field configurations that
include octupole components, even though the surface field strength is
larger (see also Gregory et al. 2008). The truncation radius is thus
given by
\be
\xi_T \approx \xi_{T0} 
\left[ 1 - {3 \over 4} \const \xi^{-2} \right]^{4/7} \approx  
\xi_{T0} \left( 1 - {3 \over 7} 
{\const \over \xi_{T0}^2} \right) \, , 
\ee
where $\xi_{T0}$ is given by equation (\ref{trundipole}), and where
the second (approximate) equality is correct only to leading order.

It is straightforward to see why the disk truncation radius becomes
smaller in the presence of an aligned octupole component: For a pure
dipole field, in a fixed poloidal plane, consider a field line that
connects the positive field region in the northern hemisphere to the
negative field region in the southern hemisphere; the field vector
${\bf B}$ points downwards in the equatorial plane of the star (the
mid-plane of the disk).  For an octupole field, however, the field
line that crosses the equator connects a negative field region in the
northern hemisphere to a positive field region in the southern
hemisphere, so the field vector ${\bf B}$ points upwards in the
equatorial plane (opposite to that of the dipole field).  For
composite field configurations, when the field vectors representing
the dipole and octupole parts of the field are added in the mid-plane,
they are anti-parallel, and the resultant field vector is less than
that of a pure dipole magnetic field. Since the field strength in the
mid-plane is smaller, the disk is able to push closer to the star.

Notice that the octupole component can be anti-parallel to the dipole.
In this case, the results of this paper remain valid, with the
parameter $\const \to -\const$. Indeed, magnetic maps of the young
star TW Hya (Donati et al. 2011b) indicate that the star has an
octupole component significantly larger than the dipole component, and
that the two contributions are anti-parallel ($\const \approx -6$).
More specifically, the negative pole of the octupole coincides with
the positive pole of the dipole and the visible rotation pole of the
star (both components are roughly aligned with respect to the rotation
axis).  For such a configuration, the field vectors of the dipole and
octupole components of the field are oriented in the same direction in
the stellar mid-plane. The field strength of the two components thus
add at the inner disk edge, so that the disk is truncated farther from
the star than it would be with a pure dipole field.

Finally, we note that --- except in extreme cases --- the dipole
component of a multipolar stellar field is dominant in determining the
disk truncation radius. The corrections due to the octupole field, as
considered here, are usually relatively small, of order ${\cal O}$
$(\const R_\ast^2/r_d^2)$. The parameter $\const$ is often observed to
lie in the range $\const = 1 - 10$; however, a stellar magnetic field
would require $\const \sim 100$ in order for the octupole component to
control disk truncation.  A quadrupole field component (not considered
here) could control truncation at a smaller relative field strength.
Notice also that for extremely large accretion rates, the truncation
radius can be much closer to the stellar radius, where the higher
order multipoles are important.  Finally, we stress that although the
dipole term tends to control $r_T$, the higher order components play a
large, even dominant, role in guiding the accretion flow.

\section{Conclusion}
\label{sec:conclude} 

This paper has re-examined the problem of magnetically controlled
accretion onto newly formed (or forming) stars. Our main results
include the following:

[1] We have constructed orthogonal coordinate systems $(p,q)$ in the
poloidal plane. One coordinate follows the magnetic field lines and
hence the accretion flow. Specifically, the variable $p$ measures the
distance along the field lines, which correspond to curves of constant
coordinate $q$. We also construct the scale factors, and hence the
differential operators, for these coordinate systems. This paper
constructs coordinate systems for magnetic field configurations with a
pure dipole component (Section \ref{sec:dipole}), for dipole and
octupole contributions (Section \ref{sec:dipoct}), and for dipole and
radial (split monopole) contributions (Appendix \ref{sec:diprad}). The
coordinates for a general multipole field can be written in terms of
Legendre polynomials, as shown in {\bf Result 6} of Appendix
\ref{sec:mathappendix} (see also Gregory 2011, Gregory et al. 2010).
This approach can thus be generalized further to incorporate ever more
complex magnetic field configurations.

[2] Steady-state, transonic accretion cannot reach free-fall
velocities if the effective equation of state is too stiff. For dipole
magnetic fields, accretion flows that appoach free-fall require
polytropic index $n > 5/2$ (see also Koldoba et al. 2002).  For a
magnetic field geometry with higher order multipoles, with highest
order given by $\ell$, this constraint is tighter and takes the
general form $n > \ell + 3/2$ (Section \ref{sec:genconstraint}).  In
particular, for the octupole fields ($\ell$ = 3) that are inferred for
many observed T Tauri star/disk systems, the index $n > 9/2$. For
cases that allow transonic free-fall flow, the pressure term becomes
negligible compared to the kinetic term in the force equation.  This
behavior is analogous to that found in the inner limit of the
generalized infall-collapse flows that form the star/disk systems
themselves (see the Appendix of Fatuzzo et al. 2004). This constraint
on the polytropic index $n$ places a corresponding constraint on the
allowed temperatures of the flow; accreting material cannot increase
its temperature by more than a factor of $\sim3$, and hence the flow
is close to isothermal.

[3] This formulation of the problem allows for the location of the
sonic points (when they exist) to be determined semi-analytically. For
isothermal flow, equation (\ref{matchsimple}) becomes a function of
the radial coordinate $\xi$ only, where the index $\divnum (\xi)$ of
the divergence operator and the rotation term $\roterm (\xi)$ depend
only on the geometry of the streamlines (given here by the geometry of
the magnetic field).  The matching condition at the sonic point is
given by equation (\ref{dipsonic}) for isothermal flow with a dipole
geometry, and by equations (\ref{matchsimple}, \ref{rotermdioct},
\ref{divindexdioct}) for isothermal flow with both dipole and octupole
components. For this latter configuration, the geometry of the field
is determined by the parameter $\const$, which sets the strength of
the octupole component of the field relative to that of the dipole.

[4] We have used this formulation to study accretion flow with dipole
magnetic field geometry (Section \ref{sec:dipole}) and with both
dipole and octupole components (Section \ref{sec:dipoct}). Compared
with the case of a pure dipole field, the inclusion of octupole
contributions funnels the flow onto a smaller region at higher
latitudes on the stellar surface.  The flow speeds and location of the
sonic points are largely unchanged, but the flow densities are much
larger (by a factor of $\sim\const$) as the flow approaches the star
and at the stellar surface. For systems with strong octupole
components, this enhancement increases the density $n_\ast$ at the
stellar surface by an order of magnitude, roughly from
$n_\ast\approx10^{12}$ g cm$^{-3}$ to $n_\ast\approx{10}^{13}$ g
cm$^{-3}$ (see Section 5.3). 

[5] The inclusion of higher order multipoles alters the predicted
location of the magnetic truncation radius $r_T$ (Section
\ref{sec:truncate}). Aligned octupole components leads to a 
{\it decrease} in the truncation radius, whereas anti-aligned
octupoles increase the truncation radius.  In either case, however,
the dipole component dominates the determination of $r_T$.

One way to summarize this approach is by outlining the parameters of
the problem: The accretion flow is assumed to follow the magnetic
field lines, which are determined by the given field geometry; for the
case of joint octupole/dipole fields, for example, the geometry is
specified through the parameter $\const$.  To find solutions to the
dimensionless problem (in the isothermal limit), we must set the
dimensionless depth $b$ of the gravitational potential well and the
location $\xi_d$ of the inner disk edge. We thus have a three
dimensional parameter space $(\const,b,\xi_d)$. The rotation parameter
$\rotcon$ is specified if we assume that the corotation point
coincides with the inner disk edge; in general, this might not hold,
and $\rotcon$ represents another parameter of the dimensionless
problem. Conversion to physical parameters requires more variables to
be specified: The magnetic truncation radius $r_T$ is a function of
the stellar field strength $B_\ast$ (in addition to $\const$), stellar
mass $M_\ast$, stellar radius $R_\ast$, and the mass accretion rate
${\dot M}$.  With these quantities determined, the sound speed $a_1$
is determined for a given value of $b$ (see equation [\ref{bdefine}]).
The width $w$ of the accretion annulus must also be set; the density
scale $\rho_1$ is then defined through equation (\ref{numdense}).
 
One goal of this work was to develop coordinate systems that allow for
a semi-analytic treatment of magnetically controlled accretion flows
in complex geometries. We have applied these results to star/disk
systems with both dipole and octupole components, but this work should
be extended in a number of directions.  In terms of theoretical
development, we have focused on the case of isothermal flow and
octupole fields.  The thermodynamics of the accretion flow can be
modeled with increasing accuracy, first by using a general polytropic
equation of state, and then by including a full treatment of heating
and cooling. The field geometries should also be generalized,
including additional multipole components and cases where the various
magnetic poles, and the rotational pole of the star, are not aligned.
This latter complication breaks the axial symmetry of the problem and
thus requires considerable development. The work presented herein is
largely theoretical, so that an important step is to apply these
techniques to specific observed sources; such modeling should also
include comparison of line profiles. Finally, these techniques can be
applied to a host of additional astrophysical problems, including
accretion onto white dwarfs \citep{canalle}, accretion in neutron star
systems \citep{gho78}, accretion in black hole systems \citep{bla82},
the solar wind \citep{ban}, magnetically controlled outflows from
planets \citep{adams}, and many others.

\acknowledgments 

This paper benefited from discussions with many colleagues, especially
Daniele Galli, Lynne Hillenbrand, and Susana Lizano. This project was
initiated during a sabbatical visit by FCA to the California Institute
of Technology, and we are grateful for the generous hospitality of the
CalTech Astronomy Department. This work was supported at the
University of Michigan through the Michigan Center for Theoretical
Physics. FCA is supported by NASA through the Origins of Solar Systems
program (grant NNX11AK87G), and by NSF through the Division of Applied
Mathematics (grant DMS-0806756). SGG is supported by NASA grant
HST-GO-11616.07-A.

\appendix 

\section{Mathematical Results for Coordinate Systems} 
\label{sec:mathappendix} 

This Appendix provides a collection of formal mathematical results that 
constrain and specify the class of coordinate systems used herein. 

\noindent 
{\bf Result 1:} For the class of orthogonal coordinate systems
considered here, the following identities must hold:
\be
\nabla q = F \left[ - {1 \over \xi} {\partial p \over \partial \theta} 
\rhat + {\partial p \over \partial \xi} \thetat \right] \, , 
\qquad 
|\nabla q|^2 = F^2 |\nabla p|^2 \, , 
\label{identone} 
\ee
where 
\be
F = \xi \sin \theta = h_\phi \, , \qquad {\rm and} \qquad 
h_p = h_q h_\phi \, . 
\label{identtwo} 
\ee

\noindent
{\it Proof:} The first identity of equation (\ref{identone}) follows
from the requirement that $\nabla q$ must be perpendicular to $\nabla
p$. The second identity follows directly from the first.  Next we show
that the form of the function $F(\xi, \theta)$ is given by the third
identity (\ref{identtwo}).  From equation (\ref{identone}), the
partial derivatives of $q$ are given by
\be
{\partial q \over \partial \xi} = - {F \over \xi} 
{\partial p \over \partial \theta} \qquad {\rm and} \qquad 
{\partial q \over \partial \theta} = \xi F 
{\partial p \over \partial \xi} \, . 
\ee
For consistency, the partial derivatives $\partial^2 q/\partial \xi 
\partial \theta$ must be the same for either ordering, which implies 
\be
- {\partial \over \partial \theta} \left( {F \over \xi} 
{\partial p \over \partial \theta} \right) = 
{\partial \over \partial \xi}  \left( \xi F 
{\partial p \over \partial \xi} \right) \, . 
\ee
If we expand and use a more compact notation, this expression 
becomes 
\be
- {1 \over \xi^2} \left( F_\theta p_\theta + F p_{\theta \theta} \right) 
= F_\xi p_\xi + F p_{\xi \xi} + {1 \over \xi} F p_\xi \, . 
\ee
The requirement of a divergence-free field, $\nabla \cdot {\bf B}$ 
= 0, implies that the scalar field $p$ must obey the Laplace equation,
$\nabla^2 p$ = 0, which requires $p$ to satisfy the differential equation 
\be
p_{\xi \xi} + {2 \over \xi} p_\xi + {1 \over \xi^2} p_{\theta \theta} 
+ {\cot \theta \over \xi^2} p_\theta = 0 \, . 
\ee
If we combine the previous two equations, we find
\be
p_\theta \left( F_\theta - \cot \theta F \right) = 
\xi p_\xi \left( F - \xi F_\xi \right)  \, . 
\ee
Since this result must hold for arbitrary field configurations, and
hence for arbitrary $p_\theta$ and $p_\xi$, the differential equations
for $F$ must individually vanish. These conditions imply that $F$ = $C
\xi \sin \theta$ as claimed (where $C$ is a constant).  The final
identity $h_p = h_q h_\phi$ follows from the definitions of the scale
factors and the second identity. $\square$

We note that this choice for $F(\xi,\theta)$ is not unique. One can
always rescale the variables (e.g., $q \to A q$) or redefine the
variables (e.g., $q \to q^2$) and obtain a valid orthogonal coordinate
system. However, the relationships between the scale factors are not
invariant under such transformations.

\noindent 
{\bf Result 2:} The integral required to evaluate the rotation term 
can be written in the form 
\be
I = \int \xi \sin\theta \left( \xhat \cdot \phat \right) 
{dp \over \left| \nabla p \right|} =  
\int \roterm (\xi) \, d\xi \, , 
\label{result2} 
\ee 
where the integral $I$ arises in the integrated form of the equation of
motion (\ref{simpleforce}) and where $\roterm(\xi)$ is defined through
equation (\ref{termrotate}).

\noindent 
{\it Proof:} Using the definition of $\roterm(\xi)$, we can write 
the integrand of $I$ in the form  
\be
\xi \sin\theta \left( \xhat \cdot \phat \right) 
{1 \over \left| \nabla p \right|} = \roterm (\xi) \, \,   
{p_\xi \over \left| \nabla p \right|^2} \, , 
\label{integrand} 
\ee
where we note that $\hat{p}$ = $\epp / |\epp|$ = 
$\nabla p / |\nabla p|$. Next we write $dp$ in the form 
\be
dp = p_\xi d\xi + p_\theta d\theta = d\xi \left[ 
p_\xi + p_\theta {\partial \theta \over \partial \xi} \right] \, . 
\label{dp} 
\ee
The field lines correspond to lines of constant $q$ so that 
\be
dq = q_\xi d \xi + q_\theta d\theta = 0 \, , 
\ee
and hence 
\be
\theta_\xi = - {q_\xi \over q_\theta} 
= {p_\theta \over \xi^2 p_\xi} \, , 
\label{thetaxi} 
\ee 
where the second equality follows from the orthogonality of the
coordinates $p$ and $q$. Combining equations (\ref{dp}) and
(\ref{thetaxi}) allows us to write $dp$ in the form
\be
dp = d\xi \, \, {1 \over p_\xi} \, 
\left[ p_\xi^2 + {1 \over \xi^2} p_\theta^2 \right] =   
d\xi \, \, { \left| \nabla p \right|^2 \over p_\xi} \, . 
\label{dpfinal} 
\ee
Using form of the integrand from equation (\ref{integrand}) and the
expression for $dp$ from equation (\ref{dpfinal}), the integral $I$
takes the form
\be
I = \int \roterm (\xi) \, \,   
{p_\xi \over \left| \nabla p \right|^2} d\xi \, \, 
{ \left| \nabla p \right|^2 \over p_\xi} = 
\int \roterm (\xi) d\xi \, . 
\ee
This confirms the result of equation (\ref{result2}). $\square$

\noindent 
{\bf Result 3:} The partial derivatives are given by 
\be
{\partial \xi \over \partial p} = {p_\xi \over H} 
\qquad {\rm and} \qquad 
{\partial \theta \over \partial p} = {p_\theta \over \xi^2 H} \, , 
\label{result3} 
\ee
where we have defined 
\be
H = \left| \nabla p \right|^2 = p_\xi^2 + {1 \over \xi^2} p_\theta^2 \, .  
\label{bighdef} 
\ee

\noindent
{\it Proof:} To show the validity of this result, begin with the 
expressions for $p(\xi,\theta)$ and $q(\xi,\theta)$ and take derivatives: 
\be
{\partial p \over \partial p} = 1 = p_\xi {\partial \xi \over \partial p} 
+ p_\theta {\partial \theta \over \partial p} \, , 
\label{pqone} 
\ee
and 
\be
{\partial q \over \partial p} = 0 = q_\xi {\partial \xi \over \partial p} 
+ q_\theta {\partial \theta \over \partial p} \, . 
\ee
One can then substitute the derivatives of $p$ for those of $q$ 
(using Result 1) to obtain
\be
- {p_\theta  \over \xi^2} {\partial \xi \over \partial p} + 
p_\xi {\partial \theta \over \partial p} = 0 \, . 
\label{pqtwo} 
\ee
Equations (\ref{pqone}) and (\ref{pqtwo}) thus provide two 
equations for the two unknowns $\partial \xi/\partial p$ and 
$\partial \theta / \partial p$, which can be solved to obtain 
the stated result of equation (\ref{result3}). $\square$

\noindent 
{\bf Result 4:} The relationship between the derivative with respect 
to the coordinate $p$ and the original gradient operator is given by 
\be
{\partial \over \partial p} = H^{-1/2} \phat \cdot \nabla =
\left| \nabla p \right|^{-1} \phat \cdot \nabla \, . 
\ee

\noindent
{\it Proof:} 
This result follows from the chain rule and the above definitions: 
\be
{\partial \over \partial p} = 
{\partial \over \partial \xi} \, {\partial \xi \over \partial p} + 
{\partial \over \partial \theta} \, {\partial \theta \over \partial p} 
= {p_\xi \over H} \, {\partial \over \partial \xi} + 
{p_\theta \over \xi^2 H} \, {\partial \over \partial \theta} \, , 
\ee
where we have used Result 3. This expression can be re-arranged to 
obtain the form 
\be
{\partial \over \partial p} = H^{-1/2} \left[ 
\left( {p_\xi \over H^{1/2}} \right) \, {\partial \over \partial \xi} + 
\left( {p_\theta \over \xi H^{1/2}} \right) \, {1 \over \xi} 
{\partial \over \partial \theta} \right] \, = 
H^{-1/2} \phat \cdot \nabla \, .
\ee
$\square$

\noindent 
{\bf Result 5:} For any coordinate system in the poloidal plane, the
components of the scalar fields $p$ and $q$ can be added: Suppose that
the magnetic field has multiple components, represented by the vector
fields $\nabla p_j$, where the index $j$ labels the component.  Let
$q_j$ be a scalar field that provides the coordinate that is
orthogonal to $p_j$.  Then one can construct a complete orthogonal
coordinate system 
\be
\pcomp = \sum_j p_j \qquad {\rm and} \qquad \qcomp = \sum_j q_j \, , 
\label{complete} 
\ee
where $\nabla \pcomp$ points in the direction of the magnetic field
and where $\nabla \qcomp \cdot \nabla \pcomp$ = 0 (see also Backus
1988).

\noindent
{\it Proof:} By construction, we find 
\be
\nabla \pcomp = \sum_j \nabla p_j = 
\left( \sum_j {\partial p_j \over \partial \xi} \right) \rhat + 
{1 \over \xi} \left( \sum_j {\partial p_j \over \partial \theta} 
\right) \thetat \, , 
\ee
and 
\be
\nabla \qcomp = \sum_j \nabla q_j = 
\left( \sum_j {\partial q_j \over \partial \xi} \right) \rhat + 
{1 \over \xi} \left( \sum_j {\partial q_j \over \partial \theta} 
\right) \thetat \, . 
\ee
We can write the derivatives of the $q_j$ in terms of derivatives 
of the $p_j$, so that this second expression becomes
\be
\nabla \qcomp = F(\xi,\theta) 
\left( - {1 \over \xi} \sum_j {\partial p_j \over \partial \theta} 
\right) \rhat + {1 \over \xi} F(\xi, \theta) 
\left( \xi \sum_j {\partial p_j \over \partial \xi} \right) \thetat \, , 
\ee
where $F(\xi,\theta)$ = $\xi \sin \theta$ is the same function for all
of the components (from {\bf Result 1}). Using this latter form for
$\nabla \qcomp$, it becomes clear that $\nabla \qcomp \cdot \nabla
\pcomp$ = 0.  $\square$

\noindent 
{\bf Result 6:} For a magnetic field configuration that corresponds to
a multipole of order $\ell$, the coordinates $(p,q)$ can be written in
the form
\be 
p = - A \, \xi^{-(\ell+1)} P_\ell (\mu) \qquad {\rm and} \qquad  
q = {A \over \ell} \, \xi^{-\ell} \, \sin^2 \theta \, 
{dP_\ell \over d\mu} (\mu) \, , 
\label{pqgeneral} 
\ee 
where $A$ is a constant, $P_\ell (\mu)$ is the Legendre polynomial of
order $\ell$, and $\mu$ = $\cos \theta$.

\noindent
{\it Proof:} The form for the function $p(\xi,\theta)$ follows from
the requirement that the scalar field $p$ must satisfy Laplace's
equation and from the definitions of multipole expansions.  We can
find the form for the second scalar field $q(\xi, \theta)$ using
either of the relations defined in {\bf Result 1}. In this context,
the first of these relations takes the form 
\be
{\partial q \over \partial \xi} = - \sin \theta \, 
{\partial p \over \partial \theta} = - A \, 
\xi^{-(\ell+1)} \sin^2 \theta {d P_\ell \over d\mu} \, ,
\ee 
which integrates to the expression of equation (\ref{pqgeneral}). 
The second relation implies 
\be
{\partial q \over \partial \theta} = \xi^2 \sin \theta 
{\partial p \over \partial \xi} = A \, (\ell+1) \, \xi^{-\ell} \, 
\sin\theta \, P_\ell (\mu) \, , 
\ee
which integrates to the same form. $\square$

Note that {\bf Result 6} shows that an individual multipole component
can be written in the form given by equation (\ref{pqgeneral}), and
{\bf Result 5} shows that the field components can be added according
to equation (\ref{complete}).  These results thus provide a blueprint
to construct coordinate systems that trace magnetic field structures
of arbitrary complexity.

\section{Coordinate System for Dipole plus Radial Field} 
\label{sec:diprad} 

In this Appendix we develop a magnetic field model that includes both
dipole and radial components, where the radial field is actually a
split monopole field. For this class of problems, we only need to
consider flow --- and hence field geometry --- in one quadrant of the
poloidal plane. The distinction between a true radial field and a
split monopole is thus unimportant for purposes of constructing the
coordinate system. This type of magnetic field structure arises when
the infall-collapse flow that forms the disk drags in magnetic field
lines from the original molecular cloud core. Another motivation for
this structure is that the Solar magnetic field can be modeled using
dipole, radial, and quadrupole terms (e.g., Banaszkiewicz et al.
1998), where the first two contributions are dominant. The coordinate
system constructed here can thus be used to study flow along magnetic
field lines for the Sun and other main-sequence stars with winds.

The magnetic field under consideration takes the form
\be
{\bf B} = B_{\rm rad} \xi^{-2} \rhat + {1 \over 2} B_{\rm dip} \xi^{-3} 
\left( 2 \cos \theta \, \rhat + \sin\theta \, \thetat \right) \, , 
\label{dipradfield} 
\ee 
where $\xi$ is a dimensionless radius.  If we scale out the dipole
field strength, the relative size of the radial field is determined by
the dimensionless parameter
\be
\beta = {2 B_{\rm rad} \over B_{\rm dip}} \, . 
\ee
For Solar fields, the parameter $\beta$ must of be order unity. An
even better fit to the observed Solar magnetic field can be obtained
by using a modified split monopole term, where the cartesian
coordinate $z \to z + a_z$, and where the length scale $a_z \approx$
1.5 $R_\odot$.  This modification can be incorporated into the
coordinate system, as shown below.

For the magnetic field configuration of equation (\ref{dipradfield}),
the two scalar fields that define the coordinate system in the
poloidal plane are given by
\be
p = - {\beta \over \xi} - {\cos\theta \over \xi^2} \, 
\quad \qquad {\rm and} \qquad \quad 
q = {\sin^2 \theta \over \xi} - \beta \cos\theta \, . 
\label{diprad} 
\ee
Next we find the basis vectors 
\be
\epp = \xi^{-3} \left( \beta \xi + 2 \cos\theta \right) \, \rhat 
+ \xi^{-3} \sin\theta \, \thetat \, , 
\ee
and 
\be
\epq = - \xi^{-2} \, \sin^2\theta \, \rhat + \xi^{-2} 
\sin\theta \left( 2 \cos\theta + \beta \xi \right) \, \thetat \, , 
\ee
where $\epphi$ is the same as before.
The scale factors are then given by 
\be
h_p = \xi^{3} \left[ \left( \beta \xi + 2 \cos\theta \right)^2 + 
\sin^2\theta \right]^{-1/2} \, ,  
\ee
and
\be
h_q = {\xi^{2} \over \sin\theta} \left[ \sin^2\theta + 
\left( 2 \cos\theta + \beta \xi \right)^2 \right]^{-1/2} \, , 
\ee 
where $h_\phi$ is the same as before. 

\begin{figure} 
\figurenum{7} 
{\centerline{\epsscale{0.90} \plotone{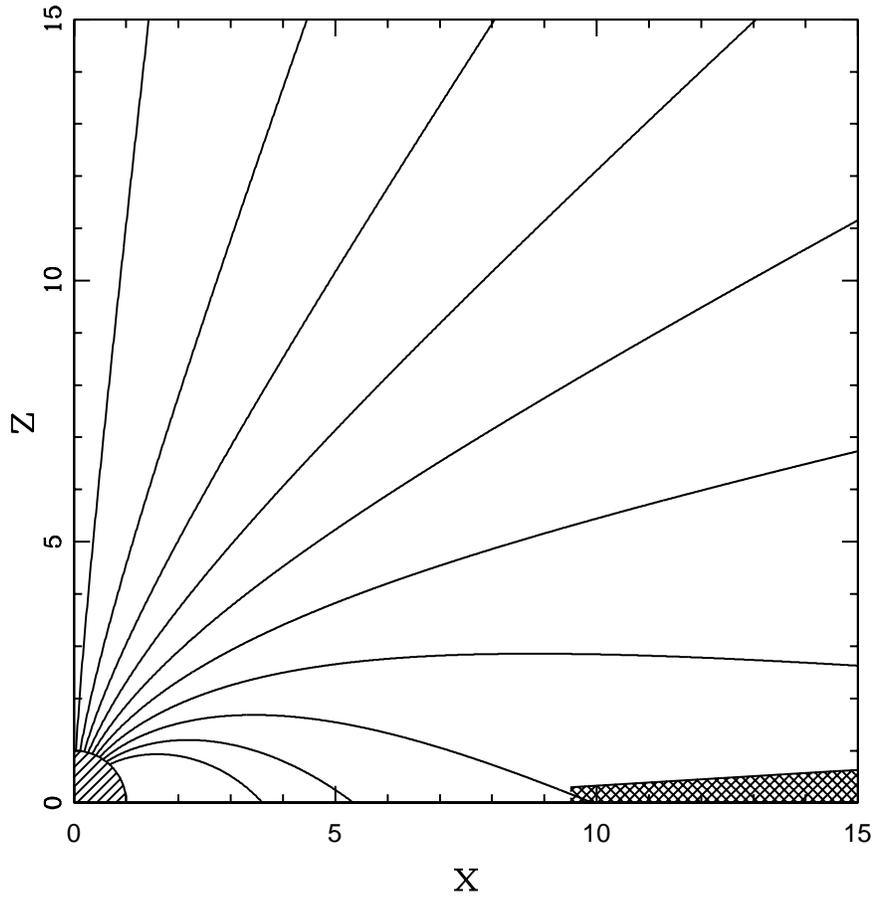} } } 
\figcaption{Magnetic field lines centered on the star for 
configurations with both dipole and split monopole contributions. The
relative strength of the monopole component is given by the parameter
$\beta$ = 1/4. The star is shown in the lower left of the figure. } 
\label{fig:drlines} 
\end{figure}

In this coordinate system, the magnetic field lines that intersect the
disk must cross the equatorial plane at dimensionless radius $\xi_d$ =
$1/q$. For a given value of $\xi_d$, the field line will hit the
stellar surface at the co-latitude given by
\be
\cos\theta_\ast = {1 \over 2} \left[ 
- \beta + \left( \beta^2 + 4 - 4/\xi_d \right)^{1/2} \right] \, . 
\ee
Notice that field lines labeled by positive $q > 0$ cross the mid-plane
and hence are closed, whereas field lines with negative $q < 0$ are
open and extend to large distances from the star.  On the stellar
surface, the angle $\theta_c$ that divides the open field lines (from
the polar regions) and the closed field lines (from equatorial
regions) is given by $2 \cos \theta_c$ = $\sqrt{\beta^2 + 4} - \beta$.
Note that the field lines that cross the equatorial plane make an
angle $A$ with respect to the vertical, where
$A=\tan^{-1}(\beta\xi_d)$; this angle must be incorporated into the
analysis (e.g., the mass accretion rate calculation of Section 3.4).
The field lines for this magnetic configuration are shown in Figure
\ref{fig:drlines}.

Along streamlines, which correspond to lines of constant $q$, we can
invert equation (\ref{diprad}) to find the angle as a function of the 
radial coordinate $\xi$, i.e., 
\be
\cos \theta = - {\beta \xi \over 2} + {1 \over 2} 
\left[ \beta^2 \xi^2 + 4 - 4 q \xi \right]^{1/2} \, . 
\ee
The field lines cross the equatorial plane, and intersect the disk, 
where $\cos\theta$ = 0, which occurs when $q \xi$ = 1. As a result, 
the field line (or streamline) that intersects the inner disk edge 
is labeled by $q = 1/\xi_d$. 
 
\begin{figure} 
\figurenum{8} 
{\centerline{\epsscale{0.90} \plotone{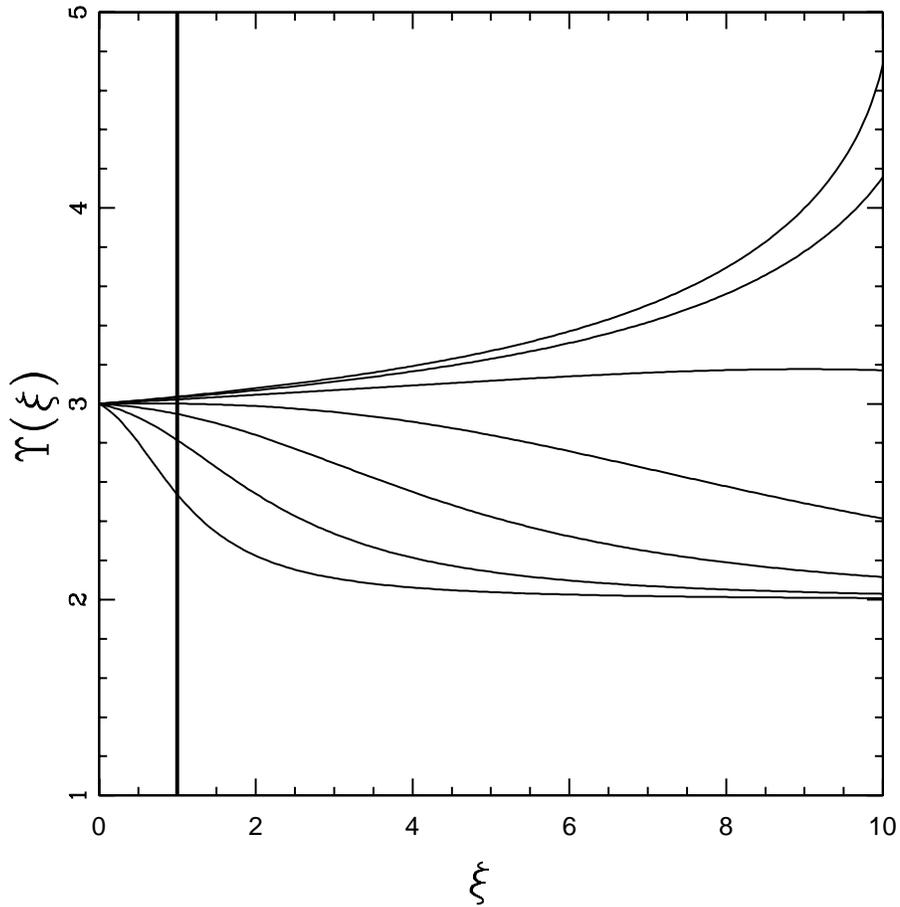} } } 
\figcaption{Index $\divnum$ of the divergence operator (see equations 
[\ref{divindex}] and [\ref{divindexdirad}]) for magnetic field
configurations with both dipole and split monopole components. The 
relative strength of the monopole component $\beta$ = 2, 1, 1/2, 1/4, 
1/8, 1/6, and 1/32 (from bottom to top). The heavy solid line at $\xi$
= 1 marks the surface of the star, where accretion flow stops.} 
\label{fig:index2} 
\end{figure}

Along a streamline, the magnetic field strength takes the form  
\be
B(\xi) = | {\bf B} | = \xi^{-3} \left\{ 4 - 3 q \xi + 
{1 \over 2} \beta^2 \xi^2 + {1 \over 2} \beta \xi
\left[ \beta^2 \xi^2 + 4 - 4 q \xi \right]^{1/2} \right\}^{1/2} \, , 
\ee
and the rotation term takes the form 
\be
\roterm (\xi) = {\xi^2 \over 4} \left\{ 2 q + \beta 
\left[ \beta^2 \xi^2 + 4 - 4 q \xi \right]^{1/2} 
- \beta^2 \xi \right\} \left\{ 3 - {\beta \xi \over 
\left[ \beta^2 \xi^2 + 4 - 4 q \xi \right]^{1/2} } \right\} \, . 
\ee
Finally, the index of the divergence operator can be written in the
form
$$
\divnum (\xi) = 3 - {\xi \over 2} \left\{ 4 - 3 q \xi + 
{1 \over 2} \beta^2 \xi^2 + {1 \over 2} \beta \xi
\left[ \beta^2 \xi^2 + 4 - 4 q \xi \right]^{1/2} \right\}^{-1} 
\, \times \, \qquad \qquad \qquad 
$$
\be 
\qquad \qquad \qquad \qquad \qquad \qquad 
\left\{ - 3 q + \beta^2 \xi + \beta 
\left[ \beta^2 \xi^2 + 4 - 4 q \xi \right]^{-1/2} 
\left[ \beta^2 \xi^2 + 2 - 3 q \xi \right] \right\}  \, . 
\label{divindexdirad} 
\ee
The index $\divnum(\xi)$ is shown in Figure \ref{fig:index2}. 

In models of the Solar magnetic field, the split monopole contribution
arises due to a current sheet that resides in a thin, disk-like
structure (e.g., Banaszkiewicz et al. 1998).  Outside the current
sheet, we can find coordinates for the magnetic field using the
methods of this paper. The relevant coordinates $(p,q)$ are given by 
\be
p = - { \cos \theta \over \xi^{2} } - {\monopole \over 
\left( \xi^2 + 2 \xi a \cos \theta + a^2 \right)^{1/2} } \, , 
\ee
and 
\be
q = {\sin^2 \theta \over \xi} - {\monopole (\xi \cos \theta + a) \over 
\left( \xi^2 + 2 \xi a \cos \theta + a^2 \right)^{1/2} } \, , 
\ee
where $a$ is a length scale that incorporates the disk-like structure
and where $\monopole$ sets the relative strength of the split monopole
contribution. In the limit $a \to 0$, and the limit $\xi \to \infty$,
we recover the simpler split monopole model of equation
(\ref{dipradfield}).  Models of the Solar magnetic field imply that
the length scale $a \approx 1.5$ (in units of the solar radius
$R_\odot$) and that the constant $\monopole \approx 0.65$.

\section{Conservation of Mass} 
\label{sec:conservemass} 

For consistency, the mass accretion rate leaving the disk surface and
that striking the stellar surface must be the same. This Appendix
shows that this constraint is satisfied. Specifically, we require that
the expression for the mass accretion rate from the disk, given by
equation (\ref{mdotdisk}), must equal that for the mass accretion rate
onto the star, given by equation (\ref{mdotstar}). The two integrals
must be identical, i.e.,
\be
\int_{\xi_d}^{\xi_2} \varpi d\varpi \lambda \, 
\left( h_p^{-1} \right)_d \, = 
\int_{\mu_1}^{\mu_2} d\mu \, 
\left( \rhat \cdot \phat \right)_\ast \, \lambda \, 
\left( h_p^{-1} \right)_\ast \, ,
\label{twoint} 
\ee
where all quantities are evaluated in the disk mid-plane for the first
integral and are evaluated at the stellar surface for the second
integral.  To show equality, we must find the relationship between the
two integration variables, those on either side of equation
(\ref{twoint}). The streamlines are given by $q = q(\xi,\theta)$,
where $q$ is a constant along a streamline. On the disk surface
$\theta$ = $\pi/2$, and scalar field $q$ is a function of the radial
variable only. We use $\varpi$ = $\xi$ to denote the radial variable
for the disk surface, and $A$ to denote the function, so that $q$ =
$A(\varpi)$ on the disk. Similarly, on the stellar surface $\xi = 1$,
and the field $q$ is a function of the angle only. Here we use the
variable $\mu$ = $\cos \theta$ to denote the variable on the stellar
surface, and $B$ to denote the function, so that $q$ = $B(\mu)$ on the
star. Since $q$ is a constant along streamlines, the relationship
between the integration variables can be derived from the identity
\be
{dA \over d\varpi} \, {d \varpi \over d \mu} = {dB \over d\mu} \, . 
\ee
Next we note that the derivative of the function $A(\varpi)$ 
takes the form 
\be
{dA \over d\varpi} = \left( \rhat \cdot \nabla q \right)_d = 
\left( - {F \over \varpi} {\partial p \over \partial \theta} 
\right)_d \, = - \varpi \left( h_p^{-1} \right)_d \, , 
\ee
where we have used the orthogonality properties of $\nabla q$ and
$\nabla p$, and where $F = \xi \sin \theta$ (see {\bf Result 1} from
Appendix \ref{sec:mathappendix}); note that $F$ = $\varpi$ on the disk
surface. Similarly, the derivative of the second function $B(\mu)$
takes the form
\be
{d B \over d \mu} = \left( - {\xi \over \sin\theta} \, \, 
\thetat \cdot \nabla q \right)_\ast = 
\left( - {\xi F \over \sin\theta} \, 
{\partial p \over \partial \xi} \right)_\ast =
- \left( \rhat \cdot \nabla p \right)_\ast \, = 
- \left( \rhat \cdot \phat \right)_\ast 
\left( h_p^{-1} \right)_\ast \, , 
\ee
where $\xi=r/R_\ast$ = 1 on the stellar surface. 
Combining the last three equations allows us to write 
\be
\varpi d \varpi \left( h_p^{-1} \right)_d  = 
\left( \rhat \cdot \phat \right)_\ast \left( h_p^{-1} \right)_\ast 
d \mu \, . 
\ee
This expression allows for a change of variables in either of the two
integrals of equation (\ref{twoint}), and thus shows that the two
integrals are equivalent.

$\,$
\bigskip 
$\,$



\begin{thebibliography}{}

\bibitem[Adams(2011)]{adams} Adams, F. C. 2011, ApJ, 730, 27

\bibitem[Adams et al.(1988)]{als88} Adams, F. C., Lada, C. J., \& Shu,
  F. H. 1988, \apj, 326, 865

\bibitem[Argiroffi et al.(2011)]{arg11} Argiroffi, C., Flaccomio, E.,
  Bouvier, J., Donati, J.-F., Getman, K. V., Gregory, S. G., Hussain,
  G. A. J., Jardine, M. M., Skelly, M. B., \& Walter, F. M. 2011,
  A\&A, 530, 1

\bibitem[Azevedo et al.(2006)]{aze06} Azevedo, R., Calvet, N.,
  Hartmann, L., Folha, D.~F.~M., Gameiro, F., \& Muzerolle, J. 2006,
  \aap, 456, 225

\bibitem[Backus(1988)]{backus} Backus, G. E., 1988, Geophysical
  Journal, 93, 413

\bibitem[Banaszkiewicz et al.(1998)]{ban} Banaszkiewicz, M., Axford,
  W. I., \& McKenzie, J. F. 1998, A\&A, 337, 940

\bibitem[Beristain et al.(2001)]{ber01} Beristain, G., Edwards, S., \&
  Kwan, J. 2001, \apj, 551, 1037

\bibitem[Blandford \& Payne(1982)]{bla82} Blandford, R.~D., \& Payne,
  D.~G. 1982, \mnras, 199, 883

\bibitem[Bouvier et al.(1993)]{bou93} Bouvier, J., Cabrit, S.,
  Fernandez, M., Martin, E.~L., \& Matthews, J.~M. 1993, \aap, 272,
  176

\bibitem[Bouvier et al.(2003)]{bou03} Bouvier, J., et al. 2003, \aap,
  409, 169

\bibitem[Bouvier et al.(2007)]{bou07} Bouvier, J., Alencar, S.~H.~P.,
  Harries, T.~J., Johns-Krull, C.~M., \& Romanova, M.~M. 2007,
  Protostars and Planets V, 479

\bibitem[Brickhouse et al.(2010)]{brick} Brickhouse, N. S., Cranmer,
  S. R., Dupree, A. K., Luna, G. J. M., \& Wolk, S.  2010, ApJ, 710, 1835

\bibitem[Calvet \& Gullbring(1998)]{cal98} Calvet, N., \& Gullbring,
  E. 1998, \apj, 509, 802

\bibitem[Canalle et al.(2005)]{canalle} Canalle, J.B.G., Saxton, C.J.,
  Wu, K., Cropper, M., \& Ramsay, G.  2005, A\&A, 440, 185

\bibitem[Donati et al.(1997)]{don97} Donati, J.-F., Semel, M., Carter,
  B.~D., Rees, D.~E., \& Collier Cameron, A. 1997, \mnras, 291, 658

\bibitem[Donati et al.(2007)]{don07} Donati, J.-F., et al. 2007,
  \mnras, 380, 1297

\bibitem[Donati et al.(2008)]{don08} Donati, J.-F., et al. 2008,
  \mnras, 386, 1234

\bibitem[Donati \& Landstreet(2009)]{don09} Donati, J.-F., \&
  Landstreet, J.~D. 2009, \araa, 47, 333

\bibitem[Donati et al.(2010b)]{don10aatau} Donati, J.-F., et
  al. 2010b, \mnras, 409, 1347 

\bibitem[Donati et al.(2010a)]{don10v2247} Donati, J.-F., et
  al. 2010a, \mnras, 402, 1426 

\bibitem[Donati et al.(2011a)]{don11v2129} Donati, J.-F., et
  al. 2011a, \mnras, 412, 2454 

\bibitem[Donati et al.(2011b)]{don11twhya} Donati, J.-F., et
  al. 2011b, \mnras.tmp.1284 [astro-ph/1106.4162] 

\bibitem[Donati et al.(2011c)]{don11v4046} Donati, J.-F., et
  al. 2011c, \mnras.tmp.1350 [astro-ph/1109.2447] 

\bibitem[Dunstone et al.(2008)]{dunst08} Dunstone, N. J., Hussain,
  G. A. J., Collier Cameron, A., Marsden, S. C., Jardine, M.,
  Stempels, H. C., Ramirez Velez, J. C., \& Donati, J.-F. 2008, MNRAS,
  387, 481

\bibitem[Edwards et al.(1994)]{edw94} Edwards, S., Hartigan, P.,
  Ghandour, L., \& Andrulis, C. 1994, \aj, 108, 1056

\bibitem[Edwards et al.(2006)]{edw06} Edwards, S., Fischer, W.,
  Hillenbrand, L., \& Kwan, J. 2006, \apj, 646, 319

\bibitem[Fatuzzo et al.(2004)]{fatuzzo} Fatuzzo, M., Adams, F., \&
  Myers, P. C. 2004, ApJ, 615, 813

\bibitem[Fischer et al.(2008)]{fis08} Fischer, W., Kwan, J., Edwards,
  S., \& Hillenbrand, L. 2008, \apj, 687, 1117

\bibitem[Ghosh \& Lamb(1978)]{gho78} Ghosh, P., \& Lamb, F.~K. 1978,
  \apjl, 223, L83

\bibitem[Ghosh \& Lamb(1979)]{ghosh} Ghosh, P., \& Lamb, F. K. 1979, ApJ, 232, 259

\bibitem[Gregory et al.(2006)]{gre06} Gregory, S.~G., Jardine, M.,
  Simpson, I., \& Donati, J.-F. 2006, \mnras, 371, 999

\bibitem[Gregory et al.(2008)]{gre08} Gregory, S.~G., Matt, S.~P.,
  Donati, J.-F., \& Jardine, M. 2008, \mnras, 389, 1839

\bibitem[Gregory et al.(2010)]{gre10} Gregory, S.~G., Jardine, M.,
  Gray, C.~G., \& Donati, J.-F. 2010, Reports on Progress in Physics,
  73, 126901

\bibitem[Gregory(2011)]{gregoryzero} Gregory, S. G. 2011,
  Am. J. Phys., 79, 461

\bibitem[Hartigan et al.(1991)]{har91} Hartigan, P., Kenyon, S.~J.,
  Hartmann, L., Strom, S.~E., Edwards, S., Welty, A.~D., \& Stauffer,
  J. 1991, \apj, 382, 617

\bibitem[Hartmann et al.(1994)]{har94} Hartmann, L., Hewett, R., \&
  Calvet, N. 1994, \apj, 426, 669

\bibitem[Herbst et al.(2007)]{herbst} Herbst, W., Eisl{\"o}ffel, J.,
  Mundt, R., \& Scholz, A. 2007, in Protostars and Planets V, eds. B.
  Reipurth, D. Jewitt, and K. Keil (Tuscon: Univ. Arizona Press), p. 297

\bibitem[Hussain et al.(2009)]{hus09} Hussain, G.~A.~J., et al. 2009,
  \mnras, 398, 189

\bibitem[Johns-Krull et al.(1999)]{joh99} Johns-Krull, C.~M., Valenti,
  J.~A., Hatzes, A.~P., \& Kanaan, A. 1999, \apjl, 510, L41

\bibitem[Johns-Krull(2007)]{joh07} Johns-Krull, C.~M. 2007, \apj,
  664, 975

\bibitem[Johnstone \& Penston(1986)]{joh86} Johnstone, R.~M., \&
  Penston, M.~V. 1986, \mnras, 219, 927

\bibitem[Kastner et al.(2002)]{kas02} Kastner, J.~H., Huenemoerder,
  D.~P., Schulz, N.~S., Canizares, C.~R., \& Weintraub, D.~A. 2002,
  \apj, 567, 434

\bibitem[Kastner et al.(2004)]{kast04} Kastner, J. H., Huenemoerder,
  D. P., Schulz, N. S., Canizares, C. R., Li, J., Weintraub, D. A.
  2004, ApJ, 605, 49

\bibitem[Kenyon \& Hartmann(1987)]{ken87} Kenyon, S.~J., \& Hartmann,
  L. 1987, \apj, 323, 714

\bibitem[Koldoba et al.(2002)]{kol02} Koldoba, A.~V., Lovelace,
  R.~V.~E., Ustyugova, G.~V., \& Romanova, M.~M. 2002, \aj, 123, 2019

\bibitem[K{\"o}nigl(1991)]{kon91} K{\"o}nigl, A. 1991, \apjl, 370,
  L39

\bibitem[Kurosawa et al.(2011)]{kur11} Kurosawa, R., Romanova, M.~M.,
  \& Harries, T.~J. 2011, \mnras, in press [astro-ph/1102.0828]

\bibitem[Li(1996)]{li96} Li, J. 1996, \apj, 456, 696

\bibitem[Li \& Wilson(1999)]{li99} Li, J., \& Wilson, G. 1999, \apj,
  527, 910

\bibitem[Lima et al.(2010)]{lim10} Lima, G.H.R.A., Alencar, S.H.P.,
  Calvet, N., Hartmann, L., \& Muzerolle, J. 2010, \aap, 522, A104

\bibitem[Long et al.(2005)]{long05} Long, M., Romanova, M.~M., \&
  Lovelace, R.V.E.  2005, ApJ, 634, 1214

\bibitem[Long et al.(2007)]{lon07} Long, M., Romanova, M.~M., \&
  Lovelace, R.V.E. 2007, \mnras, 374, 436

\bibitem[Long et al.(2008)]{lon08} Long, M., Romanova, M.~M., \&
  Lovelace, R.V.E. 2008, \mnras, 386, 1274

\bibitem[Long et al.(2011)]{lon11} Long, M., Romanova, M.~M.,
  Kulkarni, A. K., \& Donati, J.-F.  2011, \mnras, 413, 1061

\bibitem[Marsden et al.(2011)]{mars10} Marsden, S. C., Jardine, M. M.,
  Ram{\'i}rez V{\'e}lez, J. C., Alecian, E., Brown, C. J., Carter,
  B. D., Donati, J.-F., Dunstone, N., Hart, R., Semel, M., \& Waite,
  I. A. 2011, MNRAS, 413, 1922

\bibitem[Martin(1996)]{mar96} Martin, S.~C. 1996, \apj, 470, 537

\bibitem[Mohanty \& Shu(2008)]{moh08} Mohanty, S., \& Shu, F.~H. 2008,
  \apj, 687, 1323

\bibitem[Muzerolle et al.(2001)]{muz01} Muzerolle, J., Calvet, N., \&
  Hartmann, L. 2001, \apj, 550, 944

\bibitem[Najita et al.(2003)]{naj03} Najita, J., Carr, J.~S., \&
  Mathieu, R.~D. 2003, \apj, 589, 931

\bibitem[Ness et al.(2004)]{ness04} Ness, J.-U., Güdel, M., Schmitt,
  J.H.M.M., Audard, M., \& Telleschi, A. 2004, A\&A, 427, 667

\bibitem[Parker(1958)]{parkerspiral} Parker, E. N. 1958, ApJ, 128, 664

\bibitem[Parks(2004)]{parks} Parks, G. K. 2004, Physics of Space
  Plasmas : An Introduction (Boulder: Westview Press)

\bibitem[Parker(1965)]{par65} Parker, E.~N. 1965, \ssr, 4, 666

\bibitem[Radoski(1967)]{radoski} Radoski, H. R. 1967, JGR, 72, 418

\bibitem[Robitaille et al.(2007)]{rob07} Robitaille, T.~P., Whitney,
  B.~A., Indebetouw, R., \& Wood, K. 2007, \apjs, 169, 328

\bibitem[Romanova et al.(2002)]{rom02} Romanova, M.~M., Ustyugova,
  G.~V., Koldoba, A.~V., \& Lovelace, R.V.E. 2002, \apj, 578, 420 

\bibitem[Romanova et al.(2003)]{rom03} Romanova, M.~M., Ustyugova,
  G.~V., Koldoba, A.~V., Wick, J.~V., \& Lovelace, R.V.E. 2003,
  \apj, 595, 1009

\bibitem[Romanova et al.(2011)]{rom11} Romanova, M.~M., Long, M.,
  Lamb, F. K., Kulkarni, A. K., \& Donati, J.-F. 2011, \mnras, 411, 915 

\bibitem[Sacco et al.(2010)]{saccp} Sacco, G. G., Orlando, S.,
  Argiroffi, C., Maggio, A., Peres, G., Reale, F., \& Curran,
  R. L. 2010, A\&A, 522, 55

\bibitem[Saxton et al.(2007)]{saxton} Saxton, C. J., Wu, K., Canalle,
  J.B.G., Cropper, M., \& Ramsay, G. 2007, MNRAS. 379, 779 

\bibitem[Scharlemann(1978)]{scharl} Scharlemann, E. T. 1978, ApJ, 219, 617 

\bibitem[Shu(1992)]{shu92} Shu, F. H. 1992, Gas Dynamics (Mill Valley:
  Univ. Science Books)

\bibitem[Shu et al.(1994)]{shu94} Shu, F., Najita, J., Ostriker, E.,
  Wilkin, F., Ruden, S., \& Lizano, S. 1994, \apj, 429, 781

\bibitem[Skelly et al.(2010)]{ske10} Skelly, M.~B., Donati, J.-F.,
  Bouvier, J., Grankin, K.~N., Unruh, Y.~C., Artemenko, S.~A., \&
  Petrov, P. 2010, \mnras, 403, 159

\bibitem[Symington et al.(2005)]{sym05} Symington, N.~H., Harries,
  T.~J., \& Kurosawa, R. 2005, \mnras, 356, 1489

\bibitem[Ustyugova et al.(2006)]{usty} Ustyugova G. V., Koldoba,
  A. V., Romanova, M. M., \& Lovelace, R.V.E. 2006, \apj, 646, 304

\bibitem[Uzdensky(2005)]{uzd} Uzdensky, D. A. 2005, \apj, 620, 889 

\bibitem[Valenti \& Johns-Krull(2004)]{val04} Valenti, J.~A., \&
  Johns-Krull, C.~M. 2004, \apss, 292, 619

\bibitem[Waite et al.(2011)]{wait11} Waite, I. A., Marsden, S. C.,
  Carter, B. D., Hart, R., Donati, J.-F., Ram{\'i}rez V{\'e}lez,
  J. C., Semel, M., \& Dunstone, N. 2011, MNRAS, 413, 1949

\bibitem[Weinreich(1998)]{wein} Weinreich, G. 1998, Geometrical
  Vectors (Chicago: Univ. Chicago Press)

\bibitem[Yang \& Johns-Krull(2011)]{yan11} Yang, H., \& Johns-Krull,
  C.~M. 2011, \apj, 729, 83

\bibitem[Zanni \& Ferreira(2009)]{zan09} Zanni, C., \& Ferreira,
  J. 2009, \aap, 508, 1117

\end{thebibliography}
\end{document}